\documentstyle[preprint,aps,epsfig]{revtex}
\newcommand{\lsim}{\raisebox{0.3mm}{\em $\, <$}
\hspace{-3.3mm} \raisebox{-1.8mm}{\em $\sim \,$}}
\begin{document}
\draft

\title{Three flavor neutrino oscillation analysis of
the Superkamiokande atmospheric neutrino data}

\author{Osamu Yasuda\footnote{Email: yasuda@phys.metro-u.ac.jp}}
\address{Department of Physics, Tokyo Metropolitan University \\
Minami-Osawa, Hachioji, Tokyo 192-0397, Japan}

\date{April, 1998, revised August, 1998}
\preprint{
\parbox{5cm}{
TMUP-HEL-9805\\
hep-ph/9804400\\
}}

\maketitle

\begin{abstract}
Superkamiokande atmospheric neutrino data for 535 days are analyzed in
the framework of three flavor oscillations with mass hierarchy.  It
is shown that the best fit point is very close to the pure maximal
$\nu_\mu\leftrightarrow\nu_\tau$ case and $\Delta m^2 \simeq
7\times10^{-3}$ eV$^2$.  The allowed region at 90 \%CL is given and
the implications to the long baseline experiments are briefly
discussed.
\end{abstract}
\vskip 5cm

\newpage

Recent data from atmospheric neutrino experiments and especially the
Superkamiokande experiment \cite{sk1,sk2} provide very strong evidence
for neutrino oscillations.  In \cite{fvy} and \cite{gnpv} the
atmospheric neutrino data for 414 days \cite{sk1} have been analyzed
in the framework of two flavor oscillations
$\nu_\mu\leftrightarrow\nu_\tau$ or $\nu_\mu\leftrightarrow\nu_s$, and
it has been shown that both scenarios give a good fit to data.  In
this paper we extend the analysis of \cite{fvy} to the case of three
flavor oscillations using the data for 535 days \cite{sk2}. (For
quantitative three flavor analysis of atmospheric neutrinos, see
\cite{yasuda,bgk,flms,nrs,flmm,fvy3}.)  In case of general three flavor mixings
the data from the CHOOZ experiment \cite{chooz} gives a strong
constraint to any channel which involves $\nu_e$, so we include the
combined $\chi^2$ of the reactor experiments CHOOZ, Bugey \cite{bugey}
and Krasnoyarsk \cite{kra} in our $\chi^2$ analysis.

The pattern of mass squared differences with hierarchy can be
classified into two cases which are depicted in Fig. 1.
As in \cite{flms} we ignore the smaller mass squared difference,
since it is expected to be of order $10^{-5}$ eV$^2$ or
$10^{-10}$ eV$^2$ to account for the solar neutrino deficit
\cite{solar} and therefore too small to be relevant to atmospheric
neutrinos.  Furthermore we put the CP violating phase $\delta=0$ for
simplicity.  Then we are left with three parameters ($m^2,
\theta_{13}, \theta_{23}$) in the standard parametrization
\cite{pdg}.  As has been noted in
\cite{flms}, two schemes (a) and (b) in Fig. 1 are related by
exchanging the sign of $m^2$, and the case with $m^2>0$ stands for
the scheme (a) while the one with $m^2<0$ for (b) in Fig. 1.  To
evaluate the number of events, we have integrated numerically the
Schr\"odinger equation
\begin{eqnarray}
i {d \over dx} \left( \begin{array}{c} \nu_e (x) \\ \nu_{\mu}(x) \\ 
\nu_{\tau}(x)
\end{array} \right) = 
\left[ U {\rm diag} \left(0,0,{m^2 \over 2E} \right) U^{-1}
+{\rm diag} \left(0,0,A(x) \right) \right]
\left( \begin{array}{c} \nu_e (x) \\
\nu_{\mu}(x) \\ \nu_{\tau}(x)
\end{array} \right),
\label{eqn:sch}
\end{eqnarray}
where
\begin{eqnarray}
U\equiv\left(
\begin{array}{ccc}
U_{e1} & U_{e2} &  U_{e3}\\
U_{\mu 1} & U_{\mu 2} & U_{\mu 3} \\
U_{\tau 1} & U_{\tau 2} & U_{\tau 3}
\end{array}\right)
=\left(
\begin{array}{ccc}
c_{13} & 0 &   s_{13}\\
-s_{23}s_{13} & c_{23}& s_{23}c_{13}\\
-c_{23}s_{13} & -s_{23}& c_{23}c_{13}
\end{array}
\right)
\end{eqnarray}
is an orthogonal matrix, $E$ is the neutrino energy,
$A(x)\equiv\sqrt{2} G_F N_e(x)$ stands for the matter effect in the
Earth \cite{msw}.  The way to obtain the numbers of events is exactly
the same as in \cite{fvy}, and we refer to \cite{fvy} for details.  In
\cite{fvy} two quantities have been introduced to perform a
$\chi^2$ analysis.  One is the double ratio \cite{kam}
\begin{equation} 
R \equiv \frac{(N_{\mu}/N_e)|_{\rm osc}}{(N_{\mu}/N_e)|_{\rm no-osc}}
\end{equation}
where the quantities $N_{e,\mu}$ are the numbers of $e$-like and
$\mu$-like events. The numerator denotes numbers with oscillation
probability obtained by (\ref{eqn:sch}), while the denominator the
numbers expected with oscillations switched off.  The other one is the
quantity on up-down flux asymmetries for $\alpha$-like
($\alpha$=e,$\mu$) events (See also \cite{bfv,flp,fvy2,flmm}.)  and is
defined by
\begin{equation}
Y_{\alpha} \equiv {(N_{\alpha}^{-0.2}/N_{\alpha}^{+0.2})|_{\rm osc}
\over (N_{\alpha}^{-0.2}/N_{\alpha}^{+0.2})|_{\rm no-osc}},
\end{equation}
where $N_{\alpha}^{-\eta}$ denotes the number of $\alpha$-like events 
produced in
the detector with zenith angle $\cos \Theta < -\eta$, while
$N_{\alpha}^{+\eta}$ denotes the analogous quantity for $\cos \Theta >
\eta$, where $\eta$ is defined to be positive.  Superkamiokande divides the
$(-1,+1)$ interval in $\cos\Theta$ into five equal bins, so we choose
$\eta = 0.2$ in order to use all the data in the other four bins.
Thus $\chi^2$ for atmospheric neutrinos is defined by
\begin{equation}
\chi^2_{\rm atm} = \sum_E \left[\left({R^{SK} - R^{th} \over
\delta R^{SK}}\right)^2
+ \left({Y^{SK}_{\mu} - Y^{th}_{\mu} \over \delta Y^{SK}_{\mu}}\right)^2
+ \left({Y^{SK}_{e} - Y^{th}_{e} \over \delta Y^{SK}_{e}}\right)^2
\right],
\end{equation}
where the sum is over the sub-GeV and multi-GeV cases, the measured
Superkamiokande values and errors are denoted by the superscript
``SK'' and the theoretical predictions for the quantities are labeled
by ``th''.  In \cite{fvy} a $\chi^2$ analysis has been performed using
the quantities R and Y's, or using Y's only.  Throughout this paper we
use the quantities R and Y's to get narrower allowed regions for the
parameters.  We have to incorporate also the results of the reactor
experiments.  We define the following $\chi^2$:
\begin{eqnarray}
\chi^2_{\rm reactor}=
\sum_{j=1,12}^{\rm CHOOZ} \left( {x_j - y_j \over \delta x_j}
\right)^2+
\sum_{j=1,60}^{\rm Bugey} \left( {x_j - y_j \over \delta x_j}
\right)^2+
\sum_{j=1,8}^{\rm Krasnoyarsk} \left( {x_j - y_j \over
\delta x_j} \right)^2,
\end{eqnarray}
where $x_i$ are experimental values and $y_i$ are the corresponding
theoretical predictions, and the sum is over 12, 60, 8 energy bins of
data of CHOOZ \cite{chooz}, Bugey \cite{bugey} and Krasnoyarsk
\cite{kra}, respectively.  There are 6 atmospheric and 80 reactor
pieces of data in $\chi^2\equiv\chi^2_{\rm atm}+\chi^2_{\rm reactor}$
and 3 adjustable parameters, $m^2$, $\theta_{13}$ and $\theta_{23}$,
leaving 83 degrees of freedom.

Using the same parametrization as that in \cite{flms}, the results for
the allowed region of the mixing angles ($\theta_{13}$, $\theta_{23}$)
are given for various values of $m^2$ in Figs. 2 and 3.  The results
for $m^2>0$ and $m^2<0$ are almost the same.  It is remarkable that,
unlike in the case \cite{flms} of the Kamiokande data \cite{kam}, the
Superkamiokande data strongly favor $\nu_\mu\leftrightarrow\nu_\tau$
oscillations.  This is not only because we have included the combined
$\chi^2_{\rm reactor}$ of the reactor experiments but also because the
Superkamiokande data themselves favor
$\nu_\mu\leftrightarrow\nu_\tau$\cite{nb}\cite{gnpsv}.

The best fit is obtained for ($m^2, \tan^2\theta_{13},
\tan^2\theta_{23}, \chi^2) = (7\times10^{-3}$ eV$^2$,
$1.0\times10^{-2}$, 1.6, 72.8) for $m^2>0$ and ($-7\times10^{-3}$
eV$^2$, $1.0\times10^{-2}$, 1.6, 72.7) for $m^2<0$, respectively.
$\chi^2_{\rm min}=\left(\chi^2_{\rm atm}\right)_{\rm min}
+\left(\chi^2_{\rm reactor}\right)_{\rm min}=5.3+67.4=72.7$ indicates
that a fit to data is good for 83 degrees of freedom at the best fit
point.  The allowed region for $m^2$ with $\theta_{13}$, $\theta_{23}$
unconstrained is given in Fig. 4, where $\Delta\chi^2\equiv
\chi^2_{\rm atm}+\chi^2_{\rm reactor}
-\left(\chi^2_{\rm atm}+\chi^2_{\rm reactor}\right)_{\rm min}
<3.5,~6.3,~11.5$ corresponds to $1\sigma$, 90 \% CL and 99 \% CL,
respectively.  The allowed region for $|m^2|$ at 99\% CL is
$3\times10^{-4}$ eV$^2$ $\lsim |m^2| \lsim 1.8\times10^{-2}$ eV$^2$.
It should be noted that the large $m^2$ limit is excluded because
we have postulated the constraint of the reactor data\cite{largem}.

Finally, let us discuss briefly the implications of the present
analysis to the long baseline experiments \cite{k2k,minos,icarus}.
One of the interesting questions in these long baseline experiments is
whether $\nu_e$ can be observed from $\nu_\mu\leftrightarrow\nu_e$
oscillations which could be present as a fraction of the full three
flavor oscillations.  The probability $P(\nu_\mu\leftrightarrow\nu_e)$
in our scheme is given by
\begin{eqnarray}
P(\nu_\mu\leftrightarrow\nu_e)=4|U_{e3}|^2|U_{\mu 3}|^2
\sin^2\left({\Delta m^2L \over 4E}\right),
\end{eqnarray}
where $L$ stands for the path length of neutrinos.  The factor
$4|U_{e3}|^2|U_{\mu 3}|^2$ corresponds to $\sin^22\theta$ in the two
flavor framework, so by substituting this quantity in a ($\Delta m^2$,
$\sin^22\theta$) plot we can examine the possibility of observing
$\nu_e$.  The maximum values of the factor $4|U_{e3}|^2|U_{\mu 3}|^2$
in the allowed region at 90 \% CL and 99 \% CL are given in Table. 1,
respectively.  In general it is difficult for the long baseline
experiments \cite{k2k,minos,icarus} to see appearance of $\nu_e$.  In
particular, for the K2K experiment, which could probe the region of
$|m^2|$ as low as $3\times10^{-3}$ eV$^2$ for
$\nu_\mu\leftrightarrow\nu_e$ oscillations \cite{nishikawa}, it seems
very difficult to observe appearance of $\nu_e$ and disappearance of
$\nu_\mu\leftrightarrow\nu_\mu$ has to be searched for at least in the
first stage of their experiment.  On the other hand, if
$5\times10^{-4}$ eV$^2$ $\lsim |m^2| \lsim 1.0\times10^{-3}$ eV$^2$,
there is a chance for KamLAND \cite{kamland} to see a positive signal
in a disappearance experiment of
${\bar\nu}_e\leftrightarrow{\bar\nu}_e$.

In conclusion, we have analyzed the Superkamiokande atmospheric
neutrino data in the framework of the three flavor oscillations with
mass hierarchical ansatz.  We have given a allowed region at a certain
confidence level for the mass squared difference and the mixing
angles.  The data strongly favor $\nu_\mu\leftrightarrow\nu_\tau$
oscillations and therefore the most promising way in the long baseline
experiments is to search for appearance of
$\nu_\mu\rightarrow\nu_\tau$ if $\nu_\tau$ can be produced, or to look
for disappearance of $\nu_\mu\leftrightarrow\nu_\mu$ if $\nu_\tau$
cannot be produced.
\vskip 0.2in
The author would like to thank H. Minakata for discussions, R. Foot
for many useful communications and Center for Theoretical Physics,
Yale University for their hospitality during part of this work.  This
research was supported in part by a Grant-in-Aid for Scientific
Research of the Ministry of Education, Science and Culture,
\#09045036, \#10140221, \#10640280.

\newpage
\noindent
{\Large{\bf Figures}}

\begin{description}
\item[Fig.1] The hierarchical neutrino mass squared differences.
The scenarios (a) and (b) are related by exchanging
$m^2\leftrightarrow-m^2$.  They are equivalent in vacuum but
physically inequivalent in matter.

\item[Fig.2] Three flavor analysis of Superkamiokande atmospheric
neutrino data and the reactor experiments, CHOOZ, Bugey and
Krasnoyarsk.  Scenario (a) in Fig. 1 is assumed.  The solid, dashed,
dotted lines represent 68 \% CL, 90 \% CL, 99 \% CL, respectively for
degree of freedom = 3.  The right side of each panel corresponds
asymptotically to pure $\nu_\mu\leftrightarrow\nu_e$ oscillations and
the lower side to pure $\nu_\mu\leftrightarrow\nu_\tau$ oscillations

\item[Fig.3] As in Fig. 2, but the scenario (b) in Fig. 1 is
assumed.

\item[Fig.4] Value of $\Delta\chi^2\equiv
\chi^2_{\rm atm}+\chi^2_{\rm reactor}
-\left(\chi^2_{\rm atm}+\chi^2_{\rm reactor}\right)_{\rm min}
=\chi^2_{\rm atm}+\chi^2_{\rm reactor}-72.7$.  The
solid, dash-dotted, dotted lines represent the scenarios (a), (b) and
the two flavor case with maximal $\nu_\mu\leftrightarrow\nu_\tau$
mixing, respectively.

\end{description}
\newpage
\pagestyle{empty}
\vglue 3.5cm
\hglue 2.0cm
\tightenlines
\begin{tabular}{|c|c|c|}\hline
 $m^2$  & \begin{tabular}{l}
$\left( 4|U_{e3}|^2 |U_{\mu 3}|^2 \right)_{\rm max}$\\
at 90\%CL~~~~~
\end{tabular} &
\begin{tabular}{l}
$\left(4|U_{e3}|^2|U_{\mu 3}|^2\right)_{\rm max}$\\
at 99\%CL~~~~~\end{tabular}
\\\hline
$3.7\times10^{-4}$eV$^2$
& 0.00 & 0.41 \\
$4.2\times10^{-4}$eV$^2$
& 0.00 & 0.57 \\ 
$5.6\times10^{-4}$eV$^2$
& 0.54 & 0.73 \\ 
$7.5\times10^{-4}$eV$^2$
& 0.68 & 0.81 \\ 
$1.0\times10^{-3}$eV$^2$
& 0.67 & 0.82 \\ 
$1.8\times10^{-3}$eV$^2$
& 0.00 & 0.32 \\ 
$3.2\times10^{-3}$eV$^2$
& 0.07 & 0.14 \\ 
$5.6\times10^{-3}$eV$^2$
& 0.07 & 0.08 \\ 
$7.5\times10^{-3}$eV$^2$
& 0.08 & 0.10 \\ 
$1.0\times10^{-2}$eV$^2$
& 0.08 & 0.14 \\ 
$1.3\times10^{-2}$eV$^2$
& 0.00 & 0.08 \\ 
$1.8\times10^{-2}$eV$^2$
& 0.00 & 0.06 \\ 
\hline 
$-3.7\times10^{-4}$eV$^2$
& 0.00 & 0.52 \\ 
$-4.2\times10^{-4}$eV$^2$
& 0.00 & 0.60 \\ 
$-5.6\times10^{-4}$eV$^2$
& 0.57 & 0.80 \\ 
$-7.5\times10^{-4}$eV$^2$
& 0.71 & 0.84 \\ 
$-1.0\times10^{-3}$eV$^2$
& 0.68 & 0.84 \\ 
$-1.8\times10^{-3}$eV$^2$
& 0.16 & 0.35 \\ 
$-3.2\times10^{-3}$eV$^2$
& 0.08 & 0.14 \\ 
$-5.6\times10^{-3}$eV$^2$
& 0.07 & 0.09 \\ 
$-7.5\times10^{-3}$eV$^2$
& 0.08 & 0.10 \\ 
$-1.0\times10^{-2}$eV$^2$
& 0.08 & 0.14 \\ 
$-1.3\times10^{-2}$eV$^2$
& 0.00 & 0.08 \\ 
$-1.8\times10^{-2}$eV$^2$
& 0.00 & 0.06 \\\hline
\end{tabular}
\vglue 1.5cm
\begin{flushleft}
{\large Table. 1}~Maximum values of the coefficient in the probability
$P(\nu_\mu\rightarrow\nu_e)=4|U_{e3}|^2|U_{\mu 3}|^2\sin^2
\left(\Delta m^2 L/4E\right)$
allowed at a certain confidence level
of $\chi^2_{\rm atm}+\chi^2_{\rm reactor}$.
\end{flushleft}
\newpage
\pagestyle{empty}
\epsfig{file=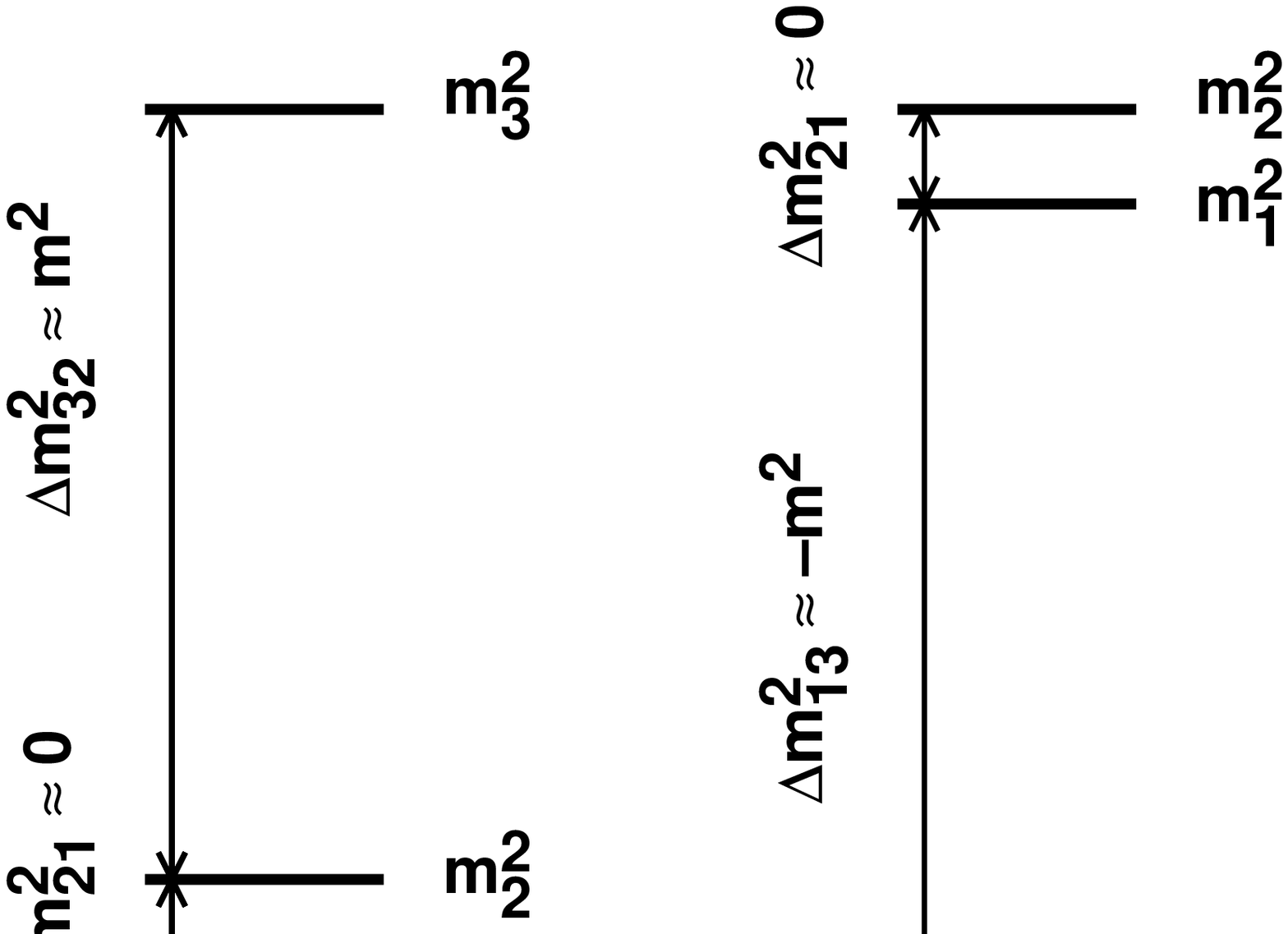,width=15cm}
\newpage
\pagestyle{empty}
\vglue 3.5cm
\hglue -1cm 
\epsfig{file=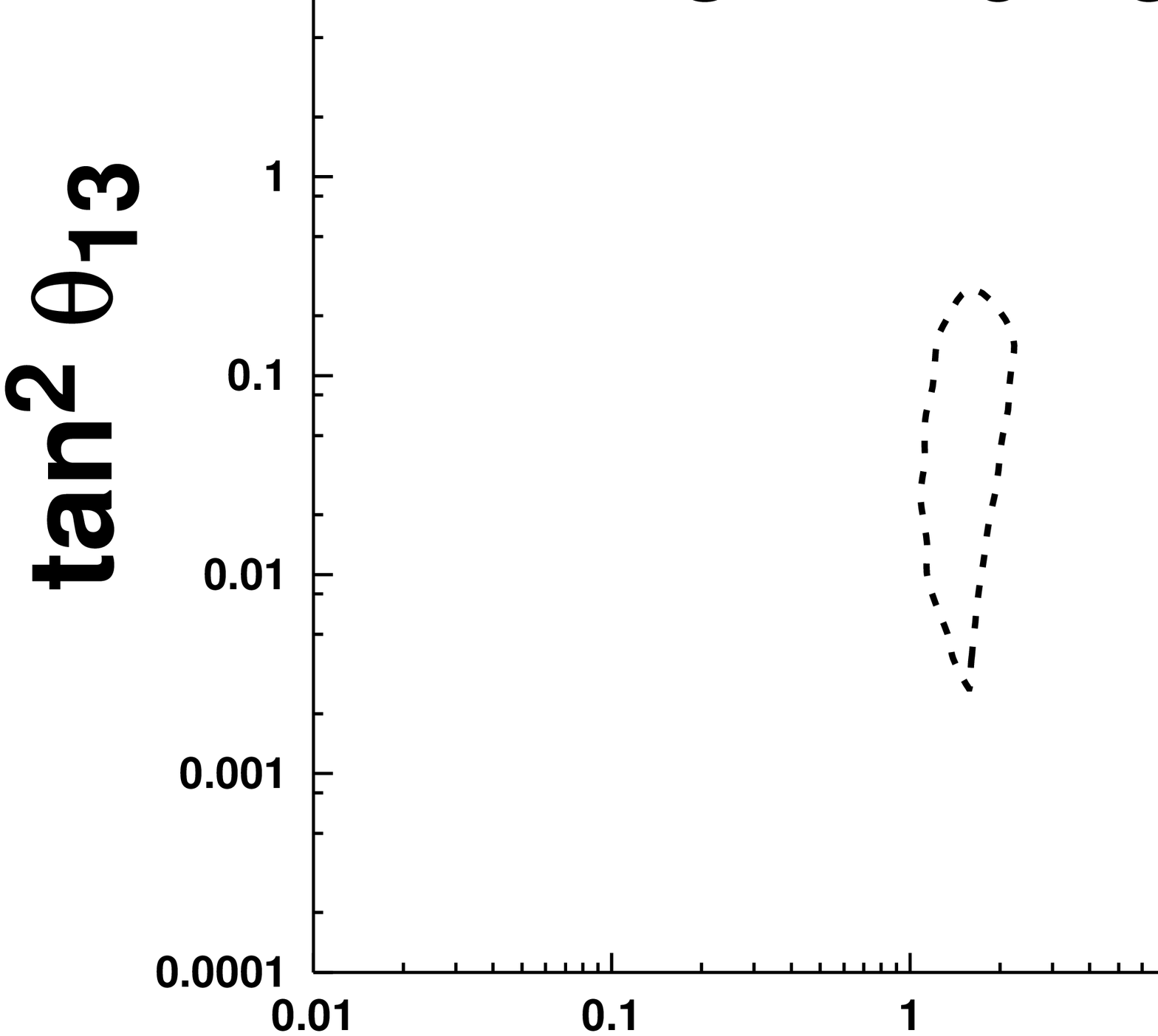,width=3.9cm}
\vglue -3.43cm \hglue 4.3cm \epsfig{file=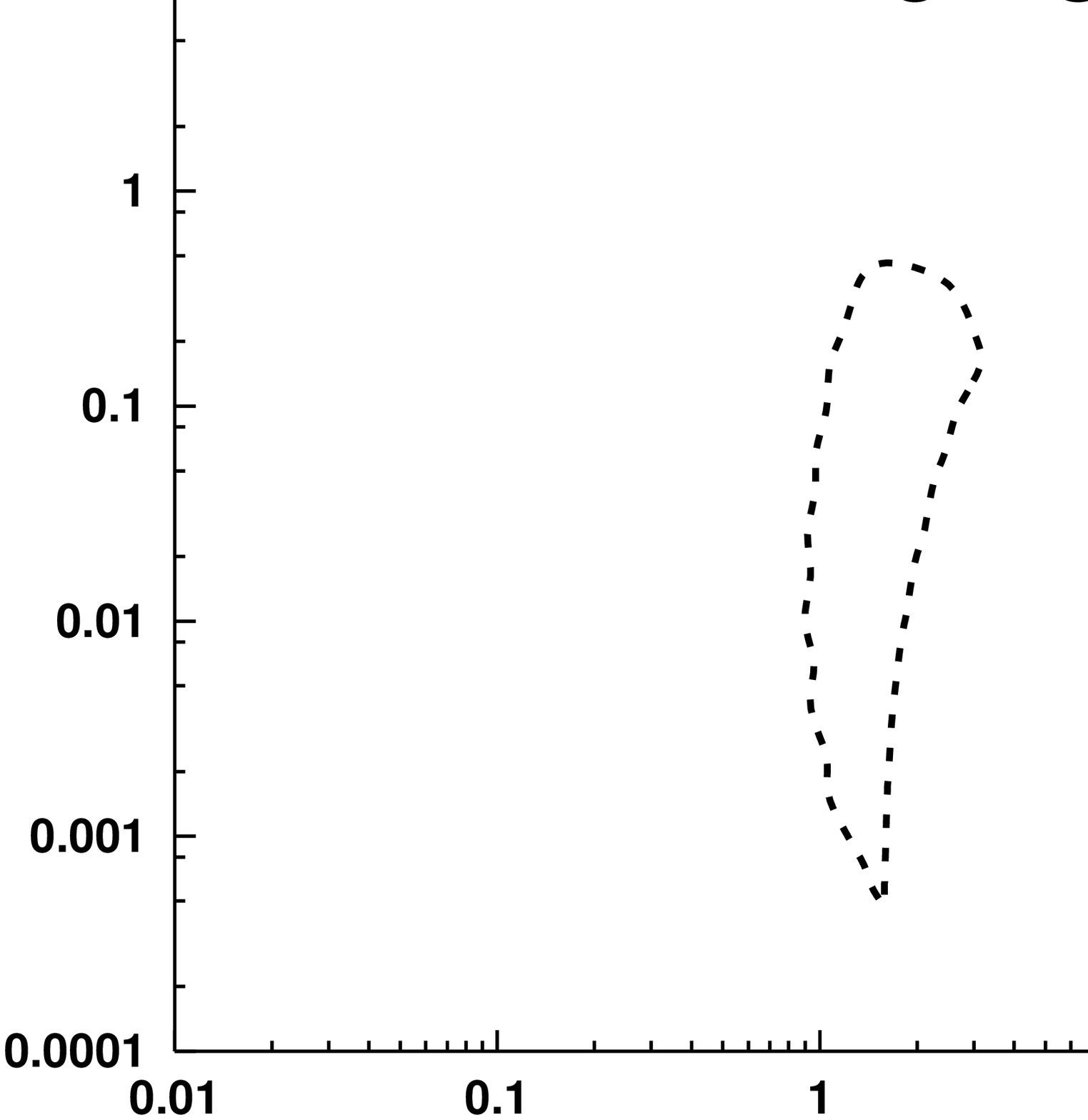,width=3.9cm}
\vglue -3.43cm \hglue 9.6cm \epsfig{file=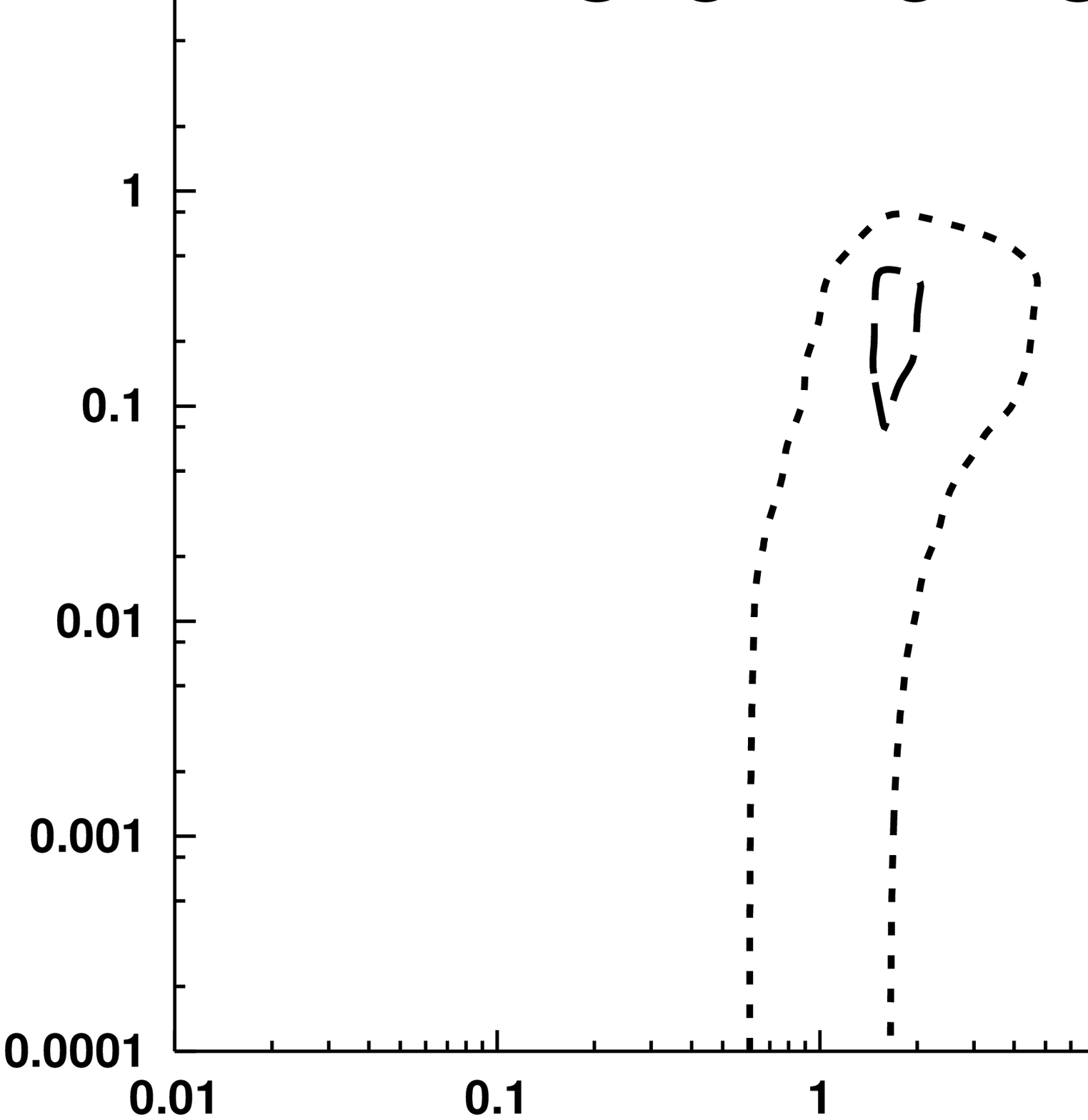,width=3.9cm}

\vglue 1.8cm
\hglue -1cm 
\epsfig{file=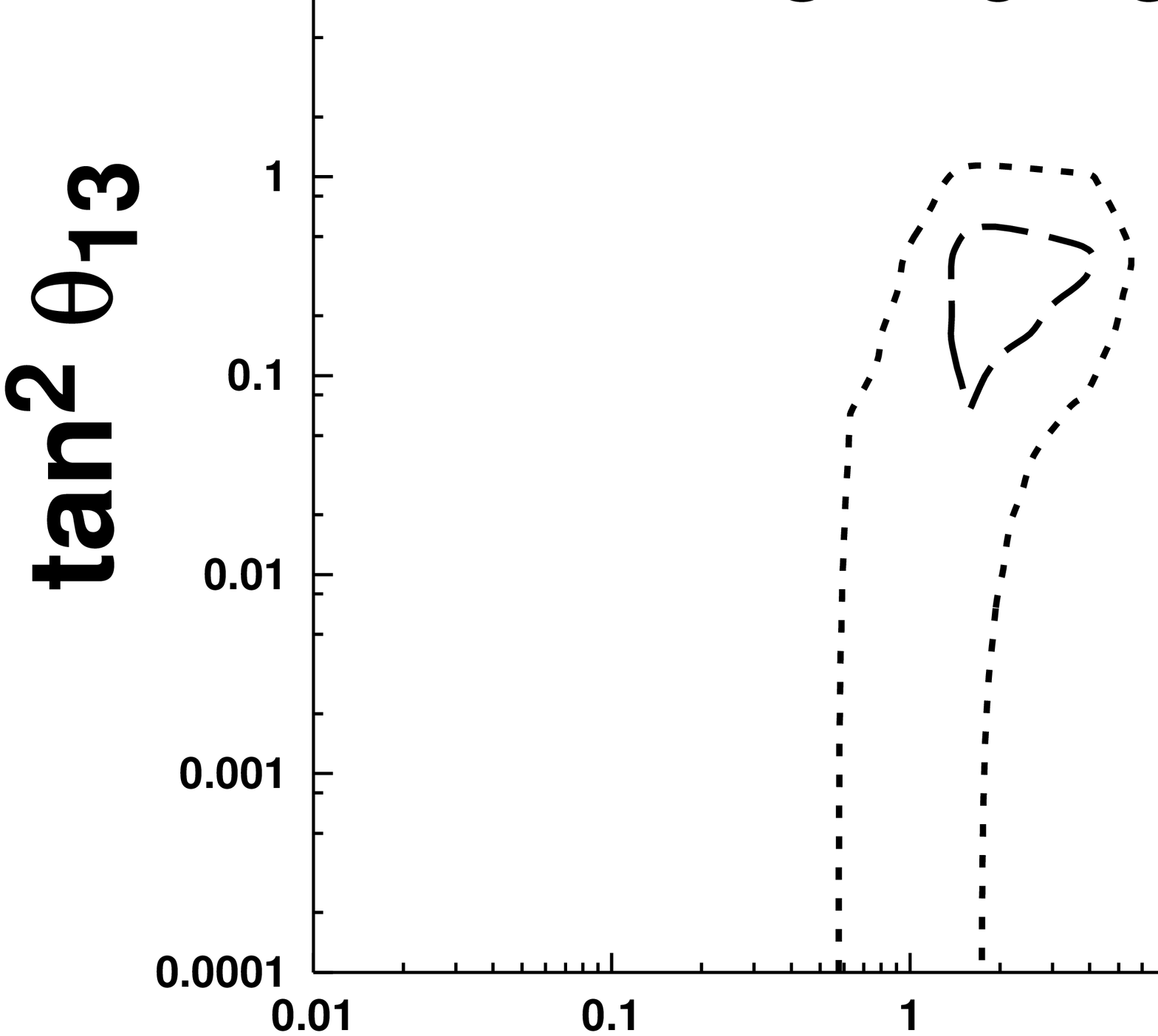,width=3.9cm}
\vglue -3.43cm \hglue 4.3cm \epsfig{file=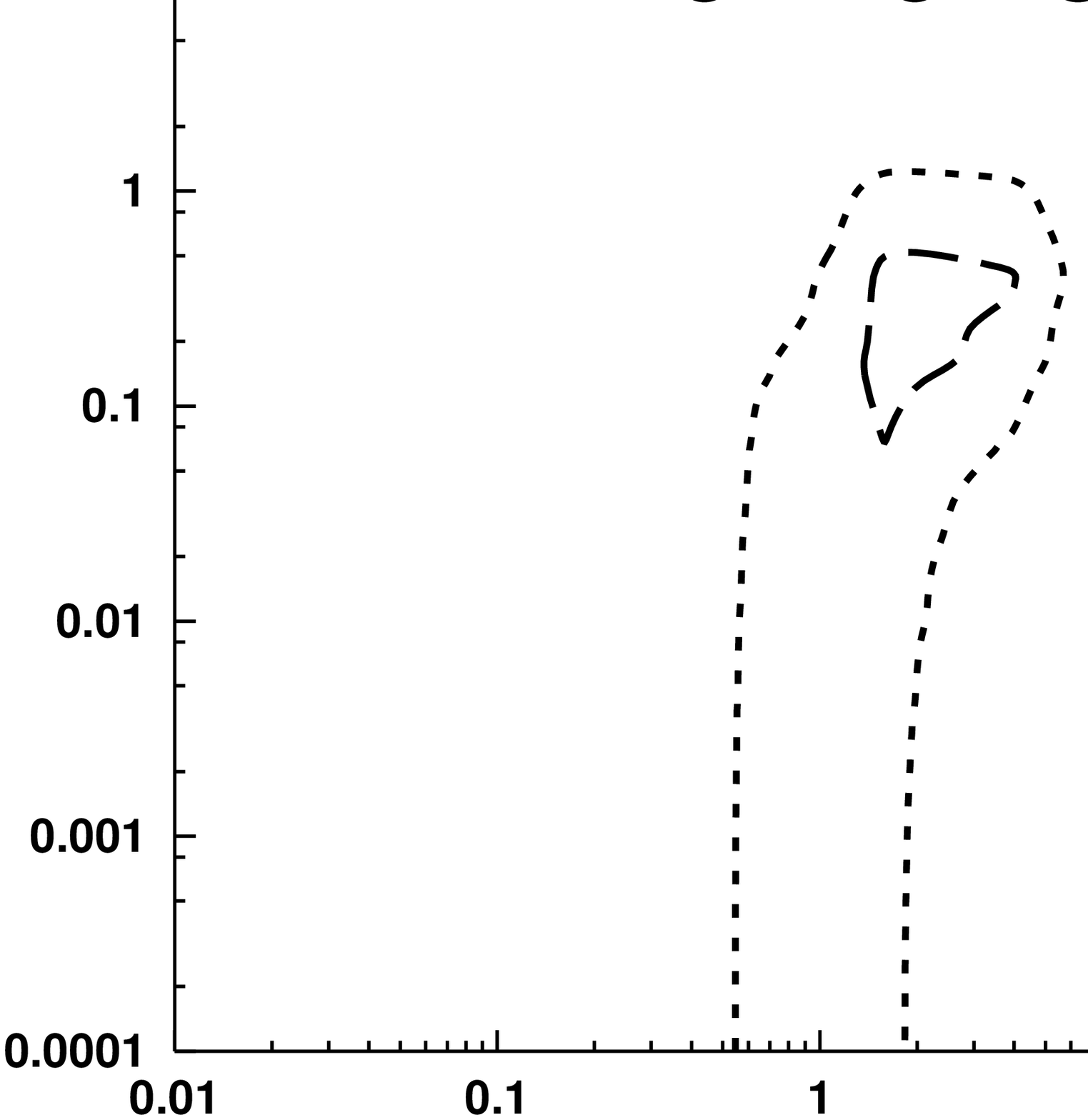,width=3.9cm}
\vglue -3.43cm \hglue 9.6cm \epsfig{file=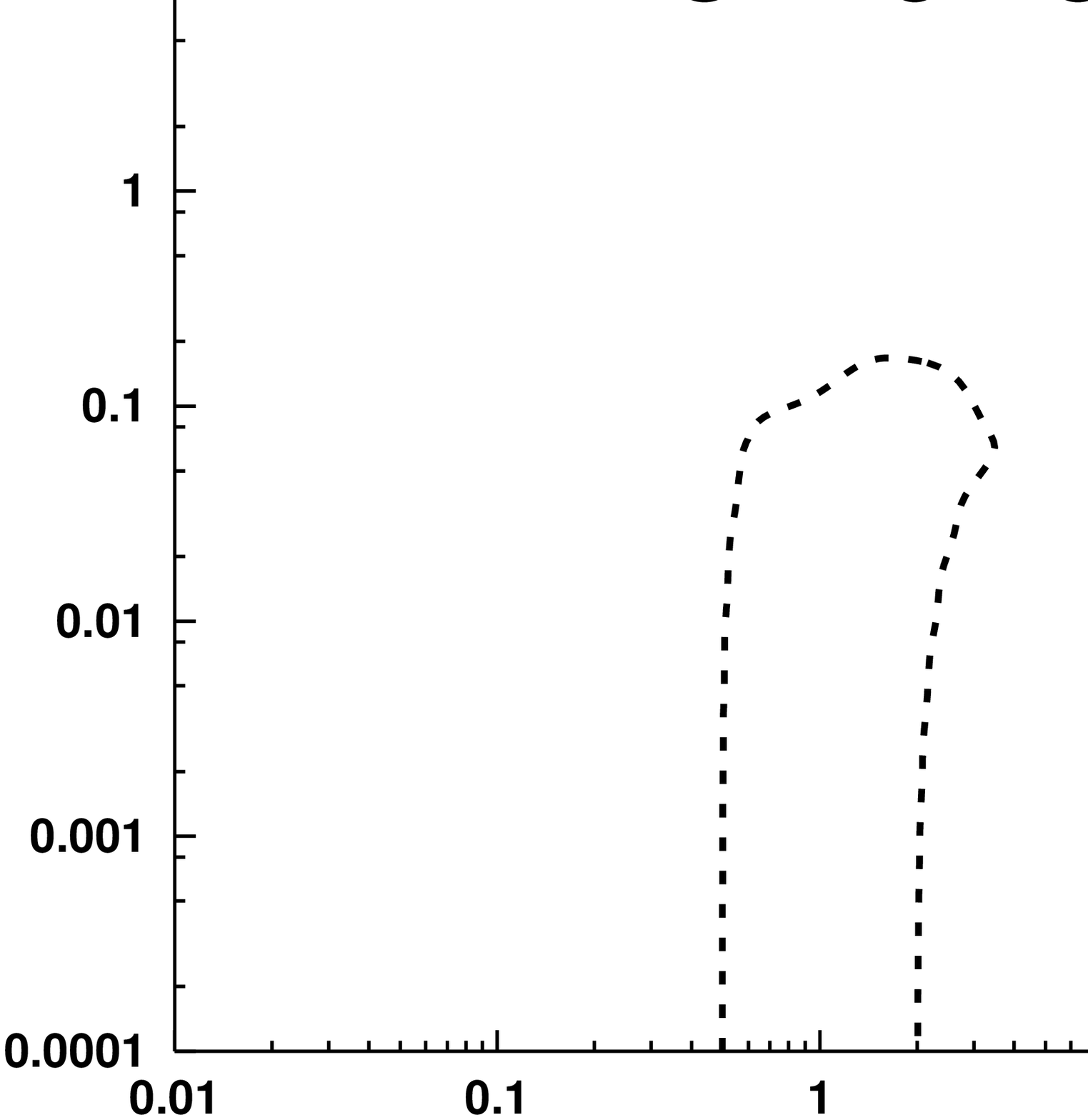,width=3.9cm}

\vglue 1.8cm
\hglue -1cm 
\epsfig{file=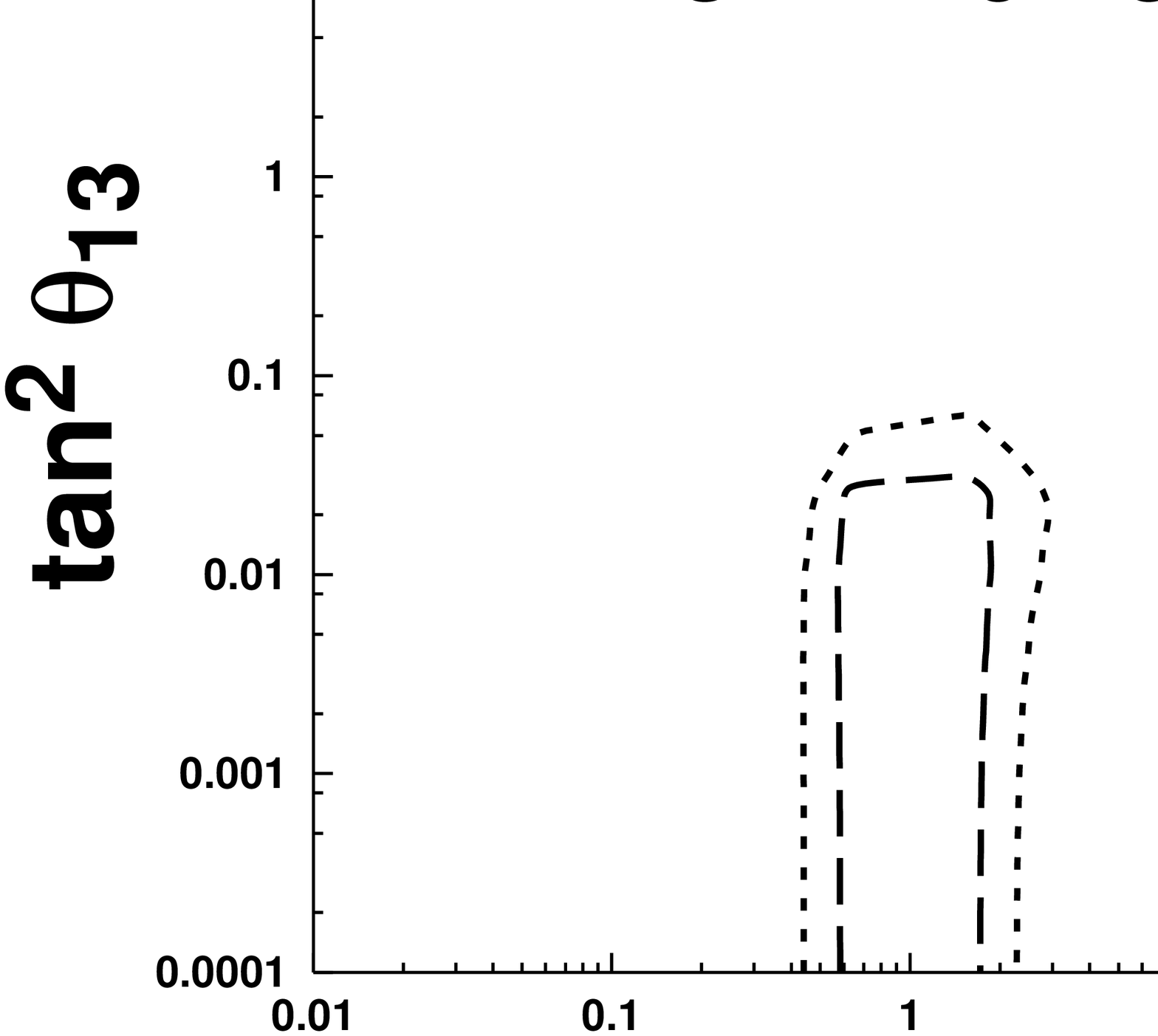,width=3.9cm}
\vglue -3.43cm \hglue 4.3cm \epsfig{file=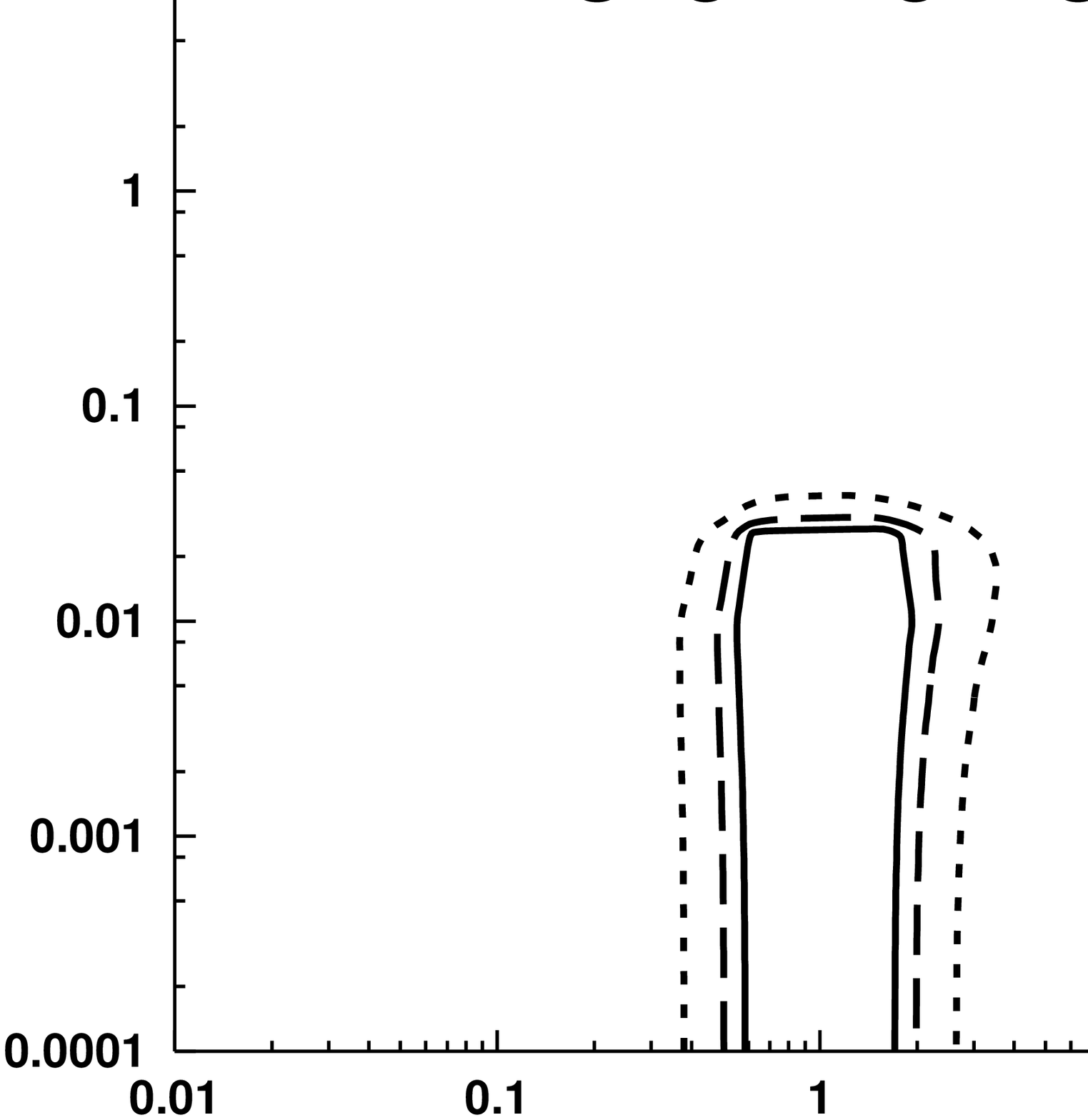,width=3.9cm}
\vglue -3.43cm \hglue 9.6cm \epsfig{file=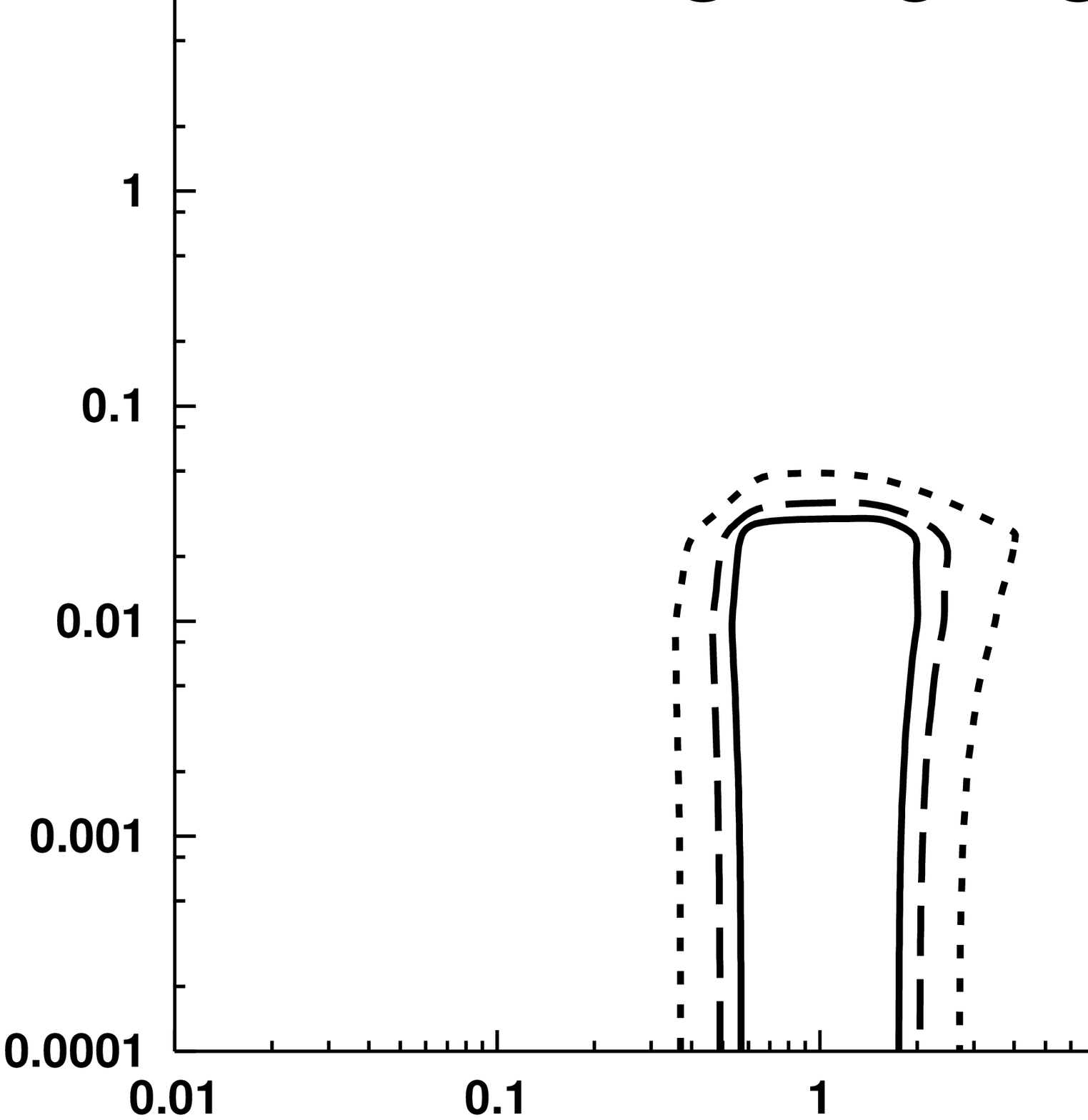,width=3.9cm}

\vglue 1.8cm
\hglue -1cm 
\epsfig{file=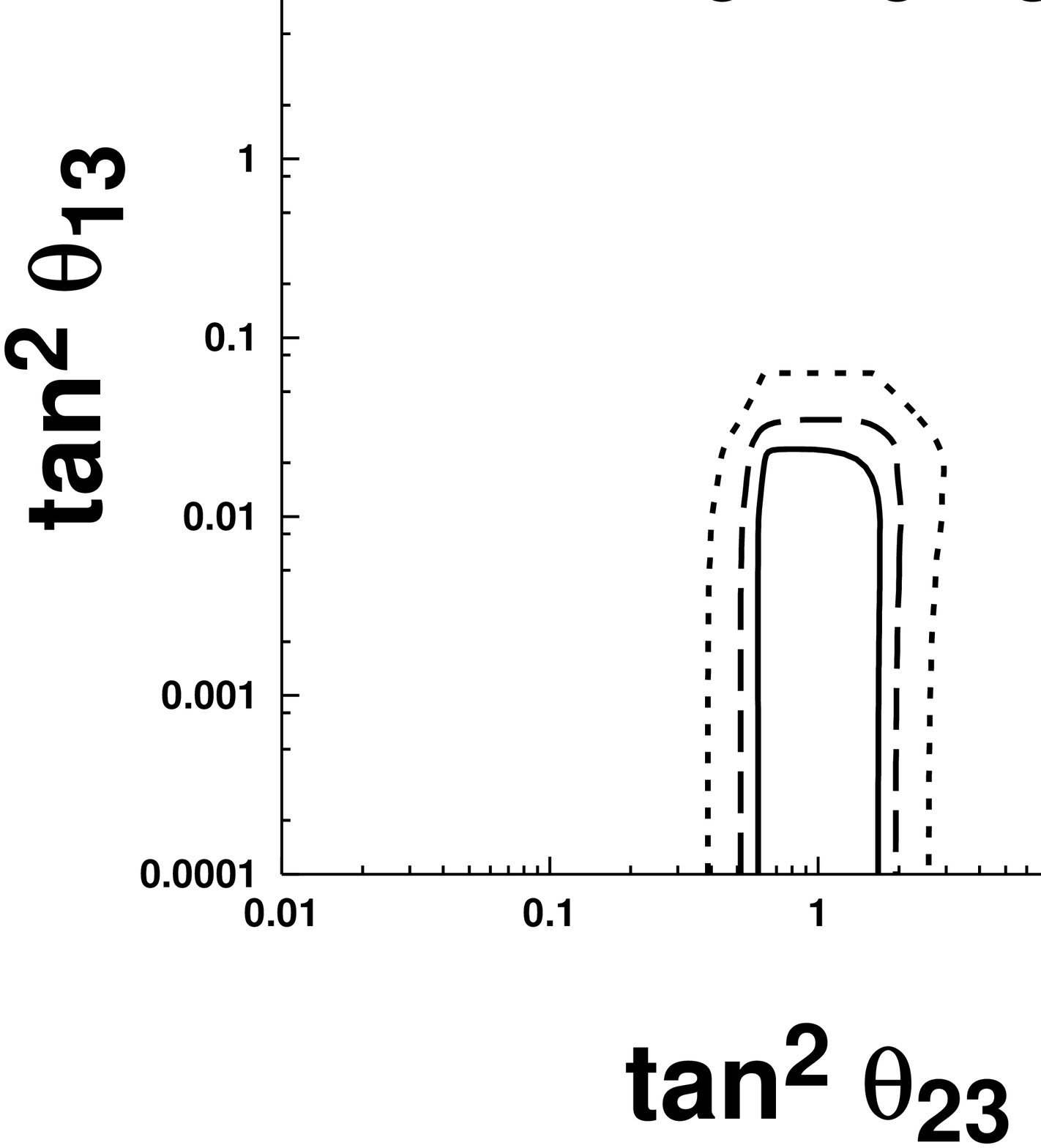,width=3.9cm}
\vglue -3.43cm \hglue 4.3cm \epsfig{file=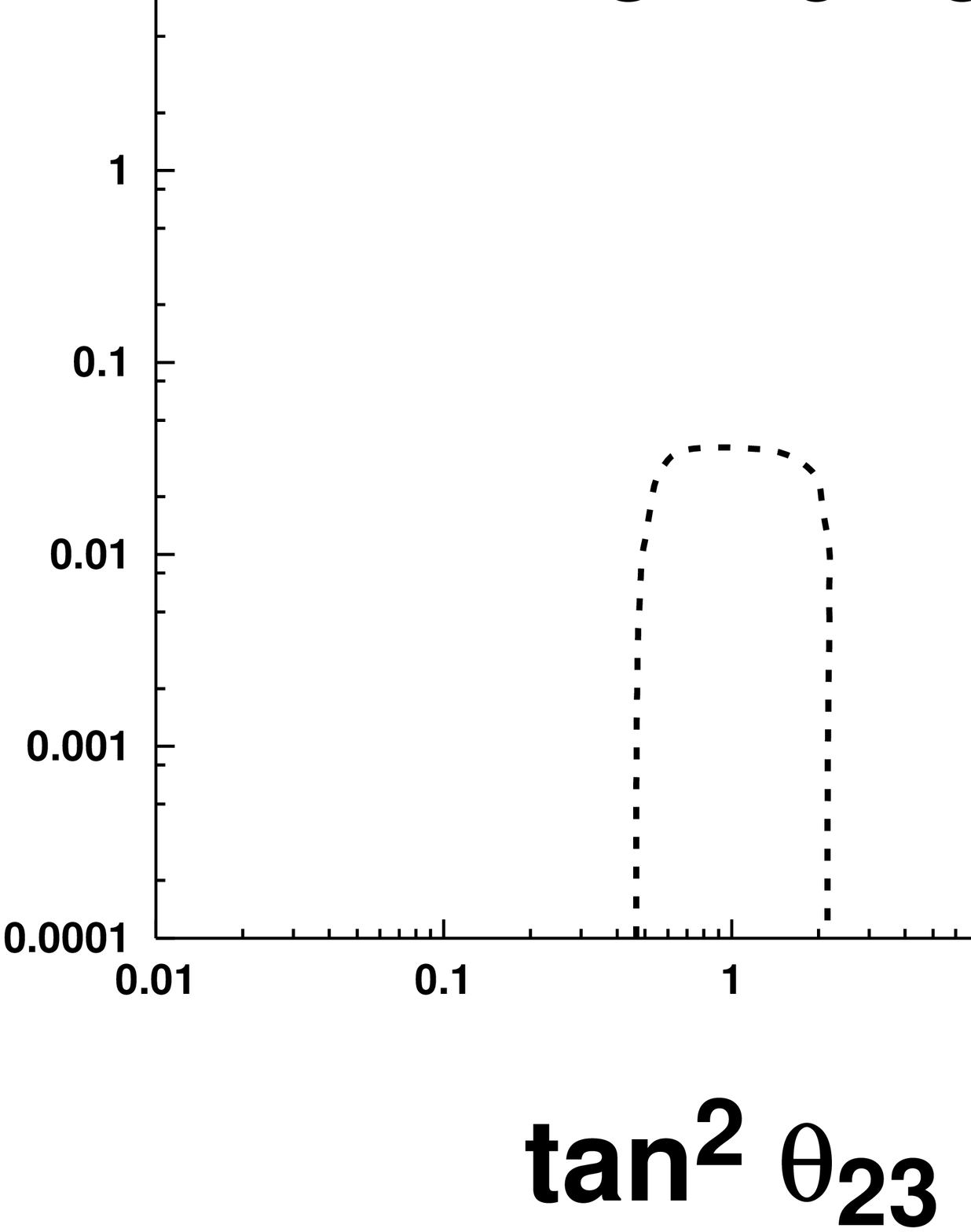,width=3.9cm}
\vglue -3.43cm \hglue 9.6cm \epsfig{file=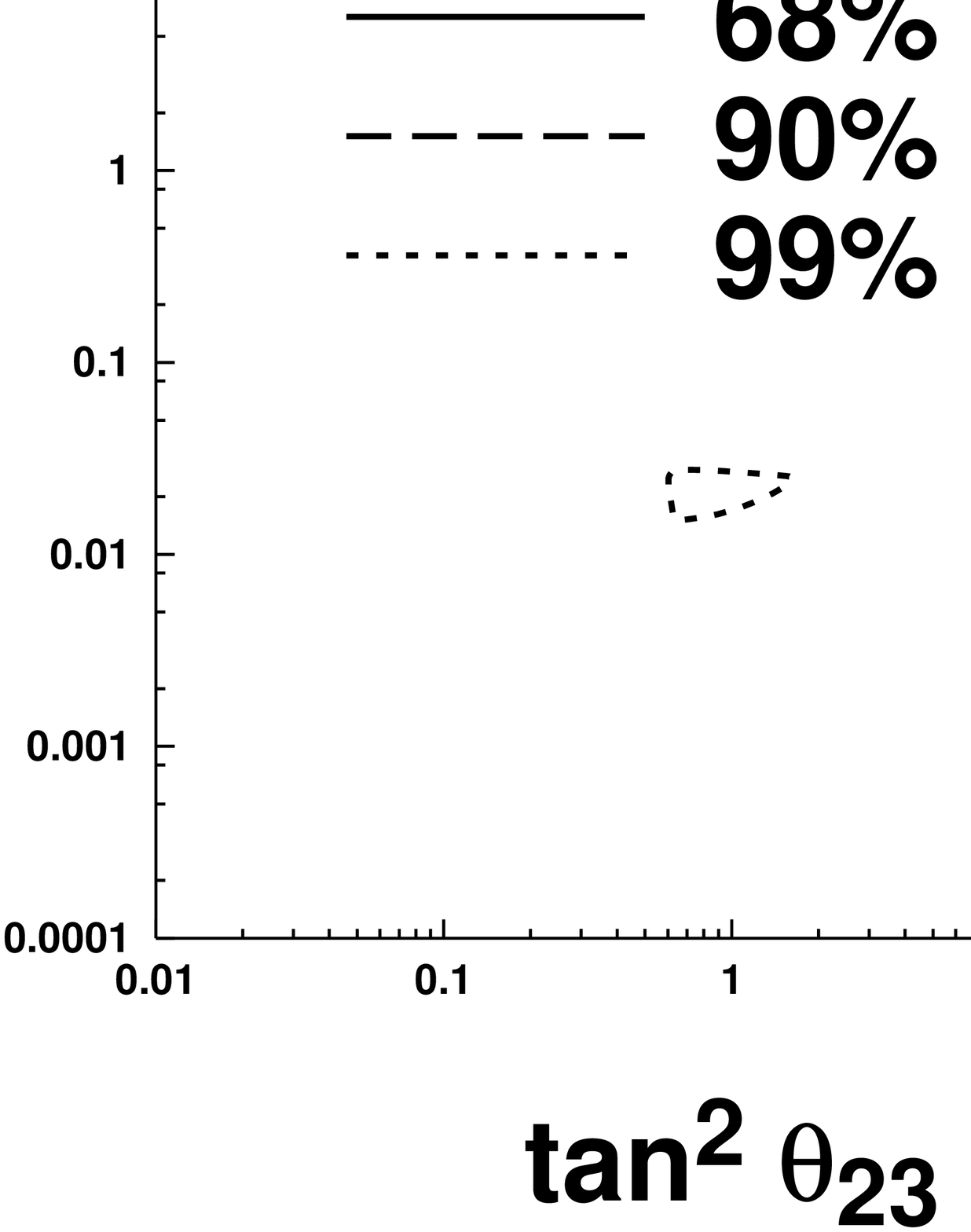,width=3.9cm}

\vglue -1.0cm
\hglue 6.0cm \epsfig{file=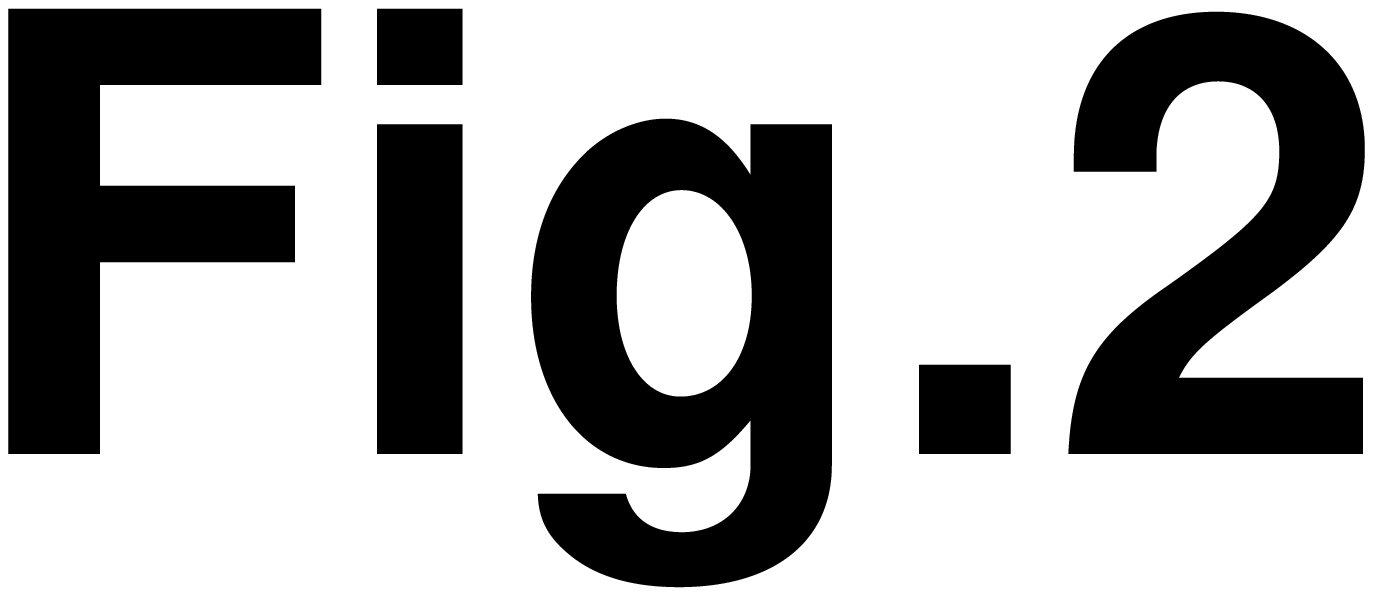,width=3.9cm}
\newpage
\pagestyle{empty}
\vglue 3.5cm
\hglue -1cm 
\epsfig{file=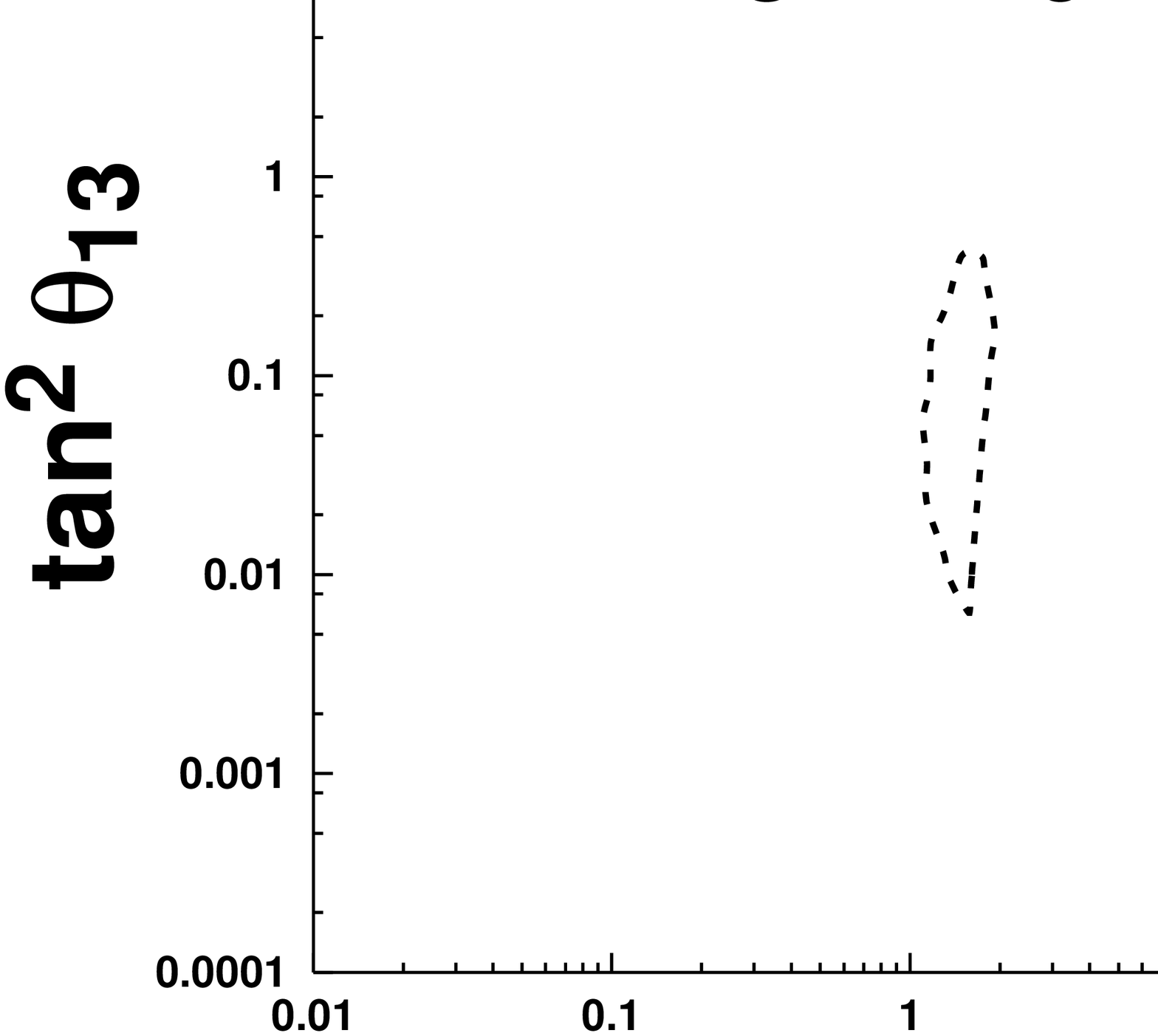,width=3.9cm}
\vglue -3.43cm \hglue 4.3cm \epsfig{file=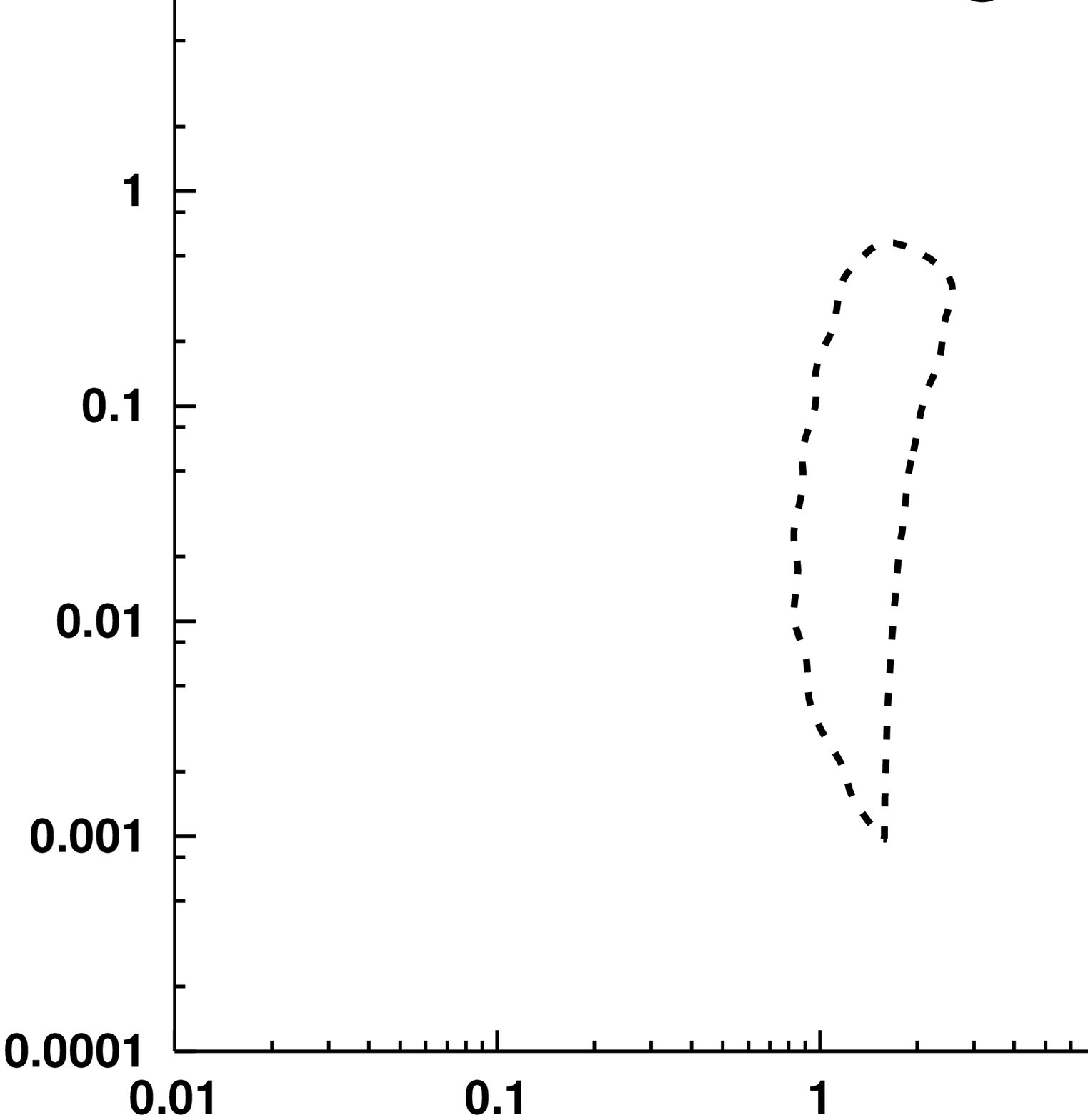,width=3.9cm}
\vglue -3.43cm \hglue 9.6cm \epsfig{file=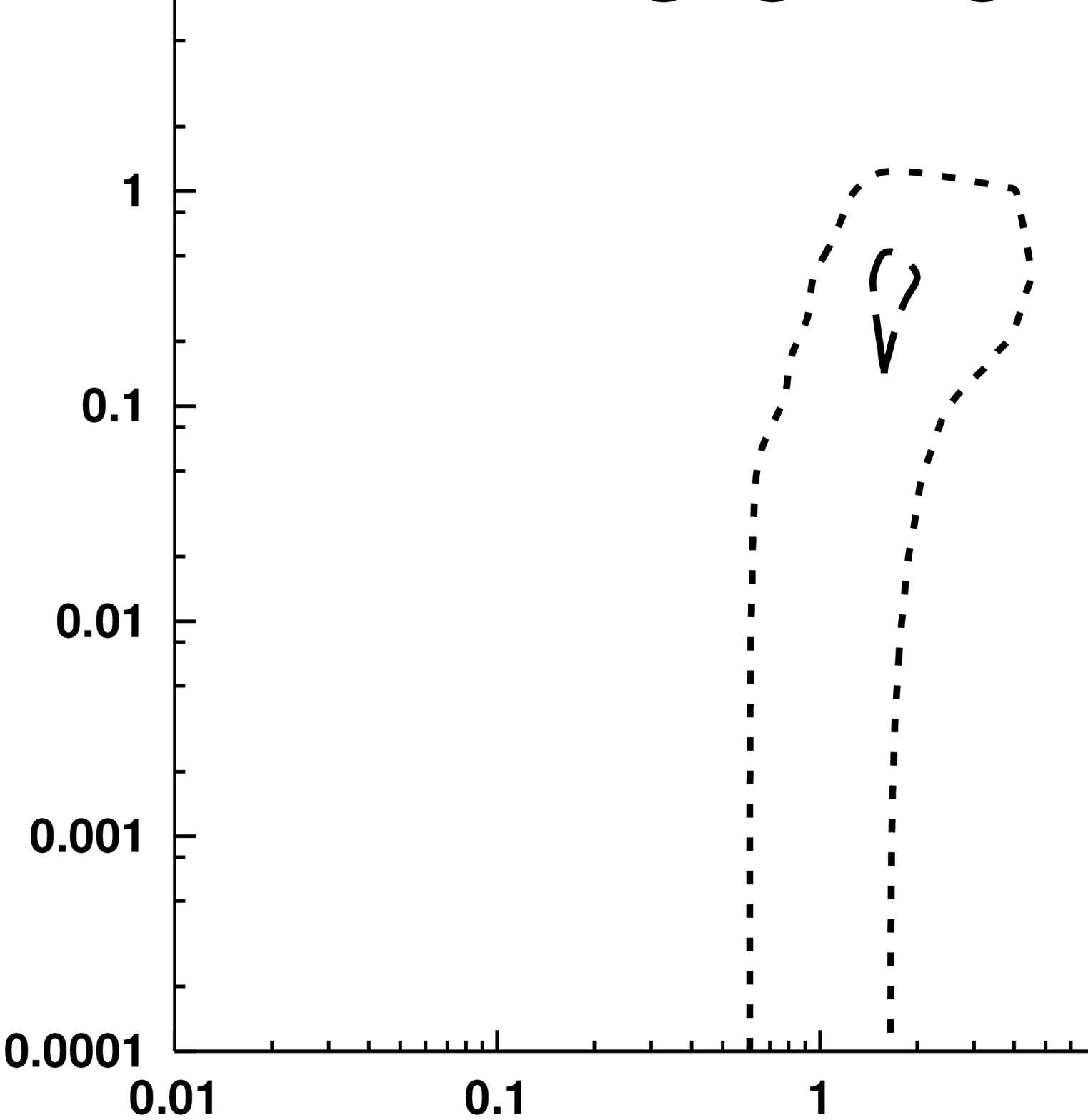,width=3.9cm}

\vglue 1.8cm
\hglue -1cm 
\epsfig{file=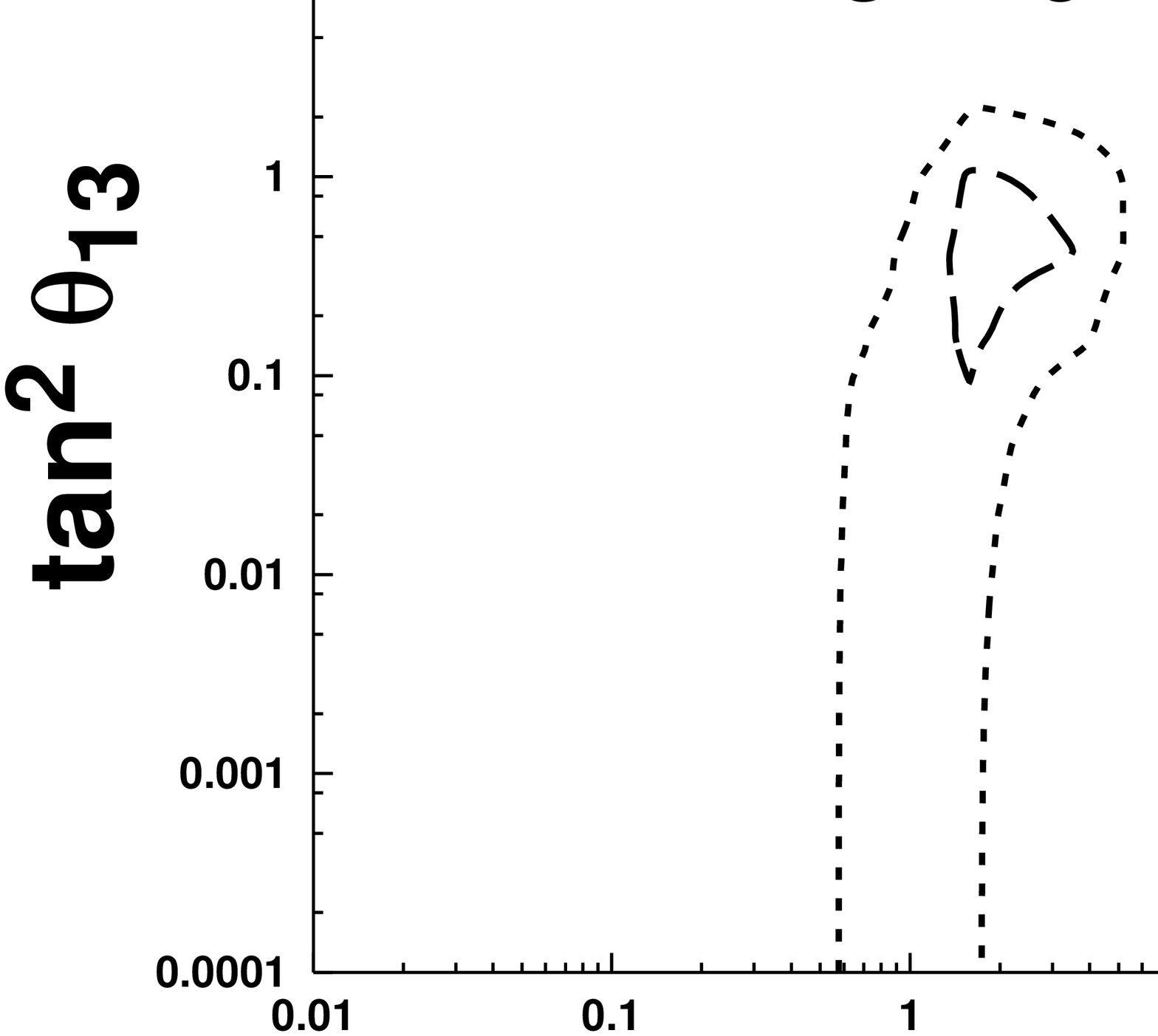,width=3.9cm}
\vglue -3.43cm \hglue 4.3cm \epsfig{file=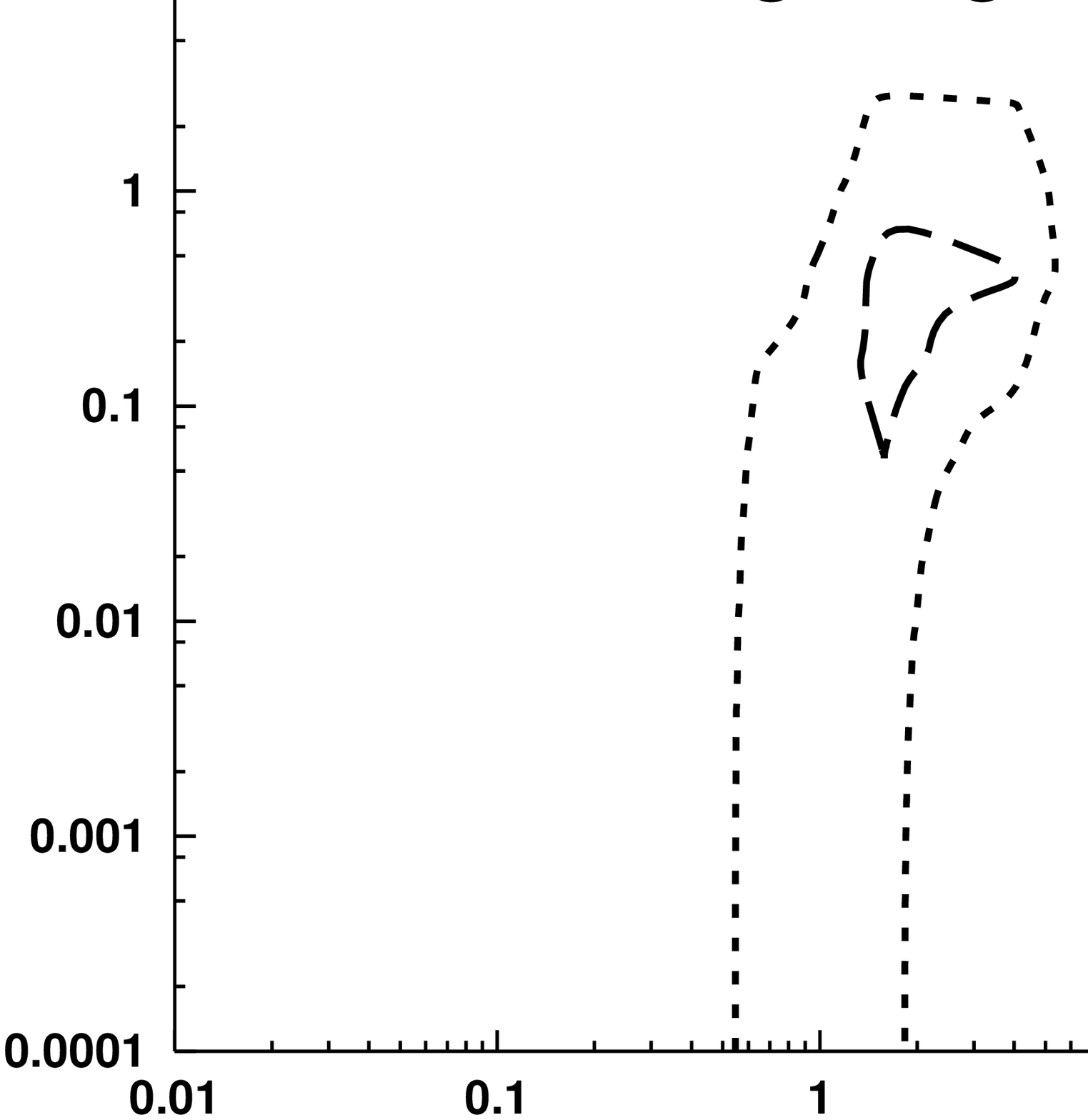,width=3.9cm}
\vglue -3.43cm \hglue 9.6cm \epsfig{file=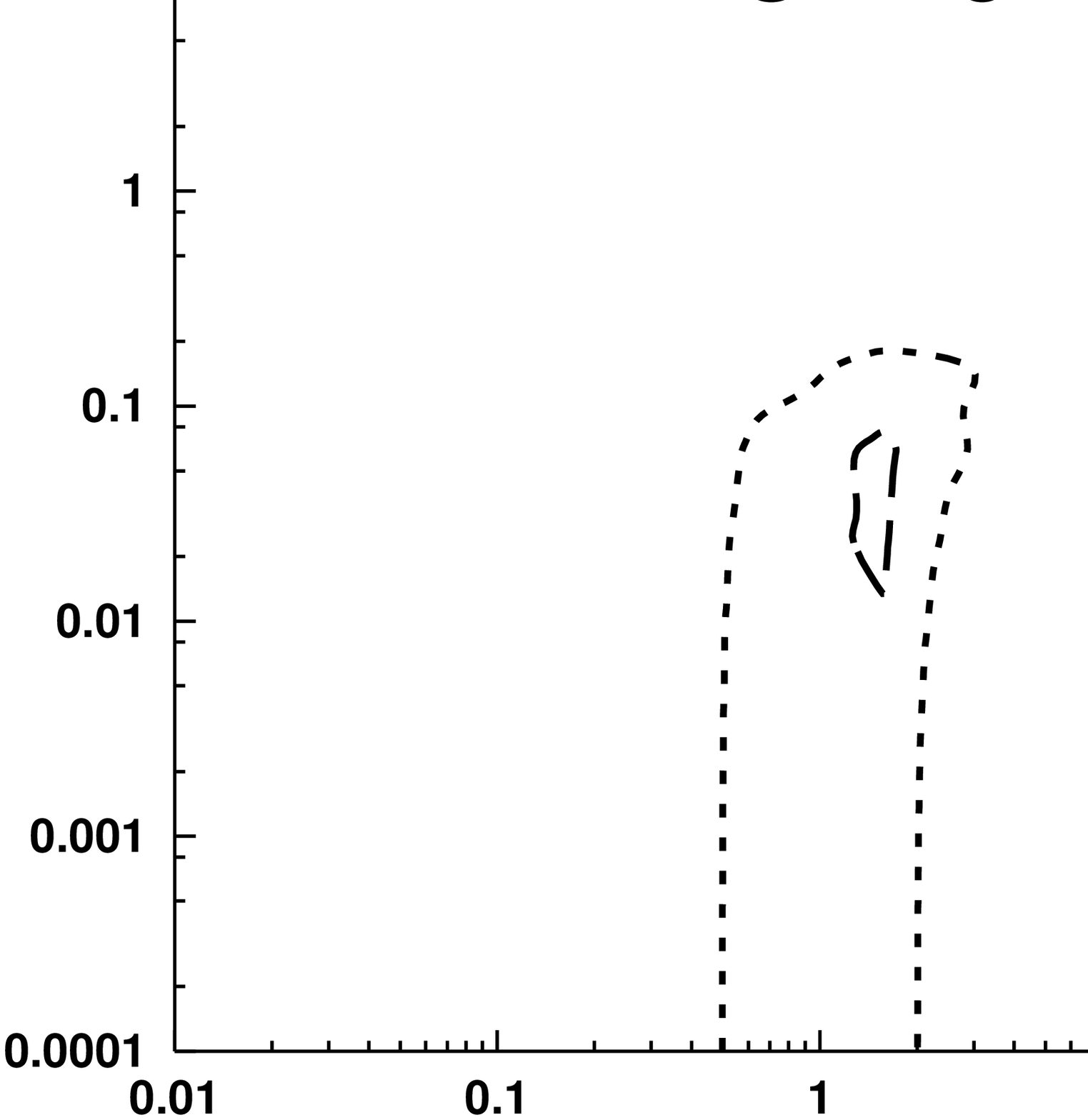,width=3.9cm}

\vglue 1.8cm
\hglue -1cm 
\epsfig{file=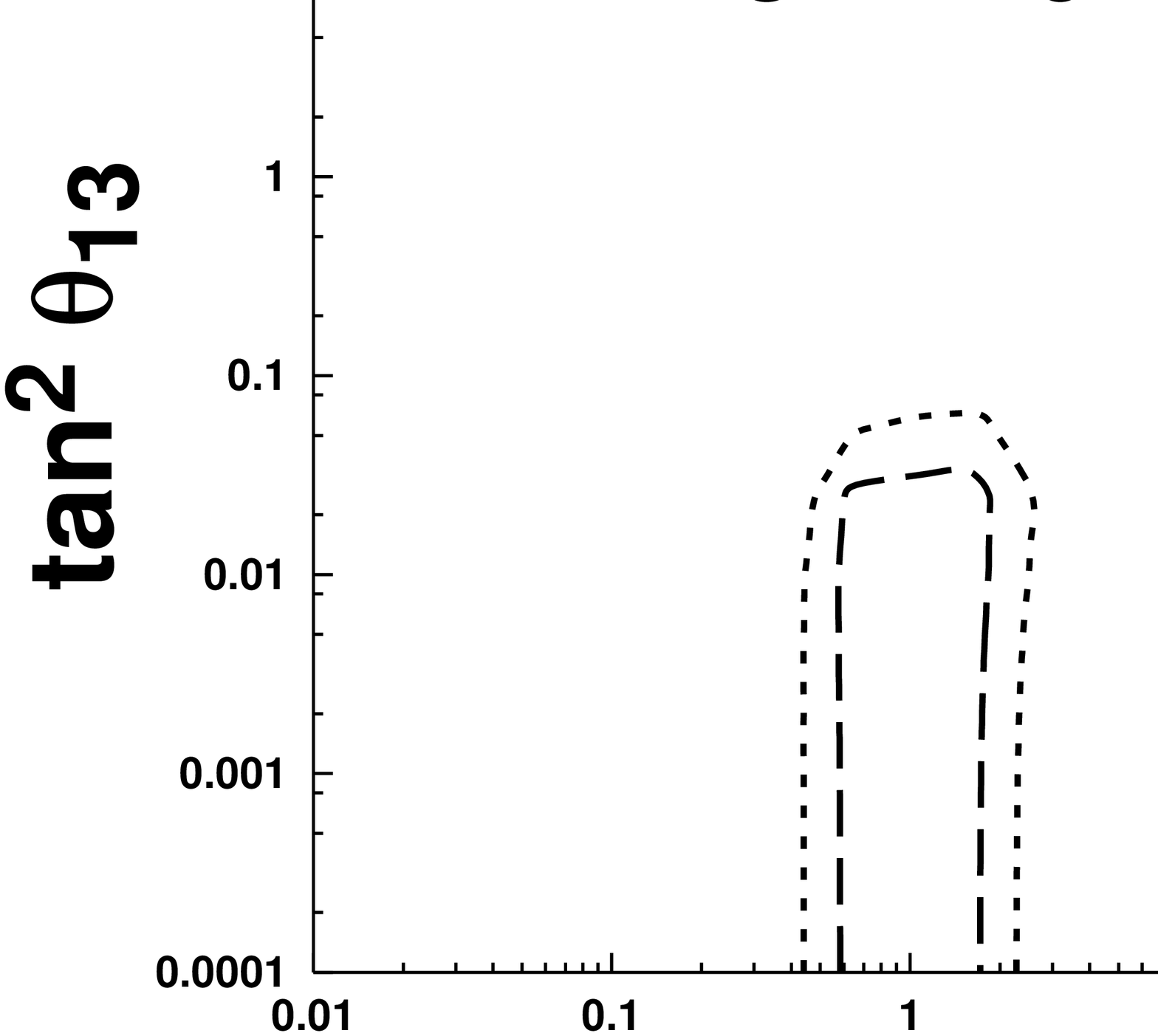,width=3.9cm}
\vglue -3.43cm \hglue 4.3cm \epsfig{file=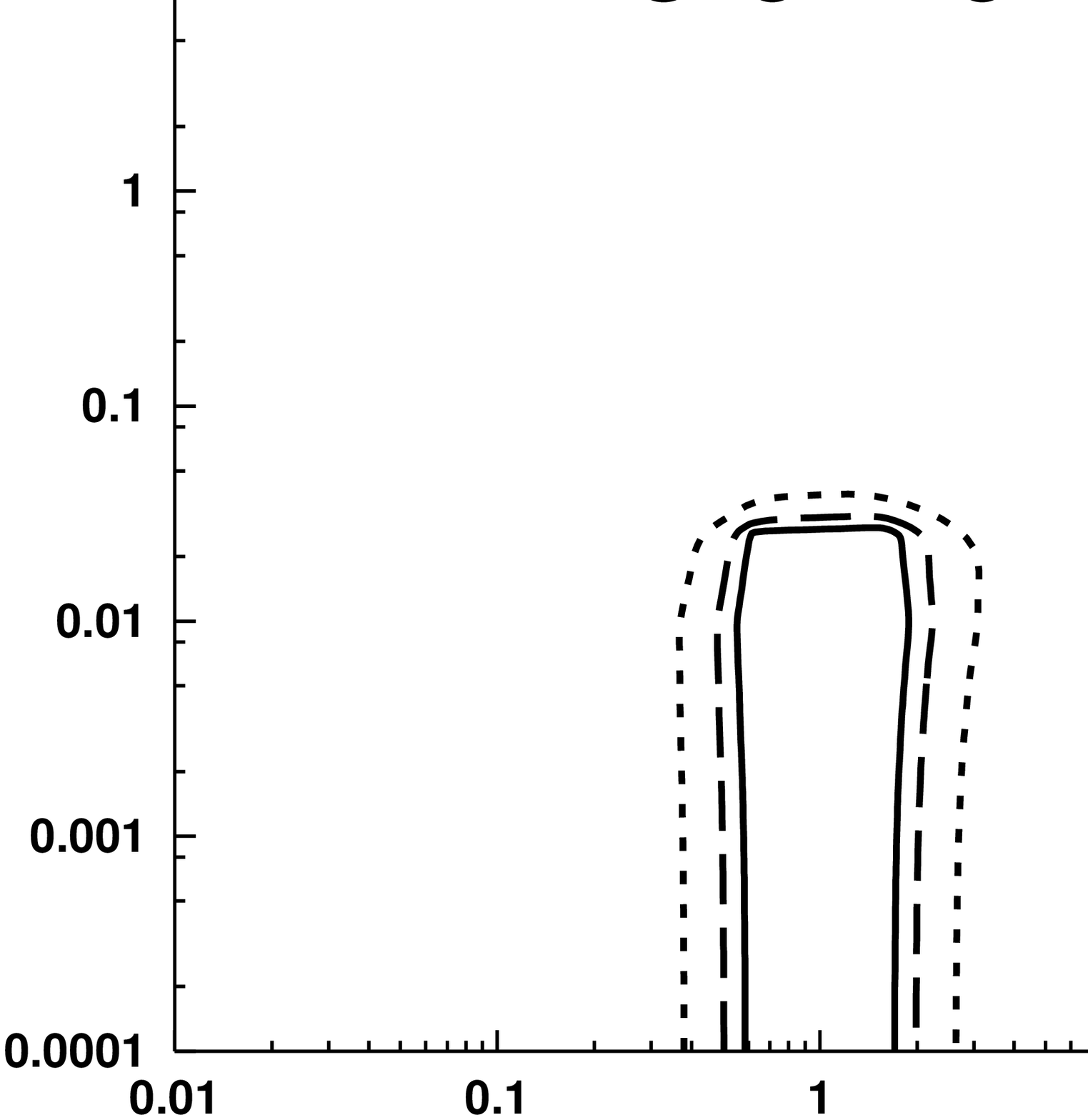,width=3.9cm}
\vglue -3.43cm \hglue 9.6cm \epsfig{file=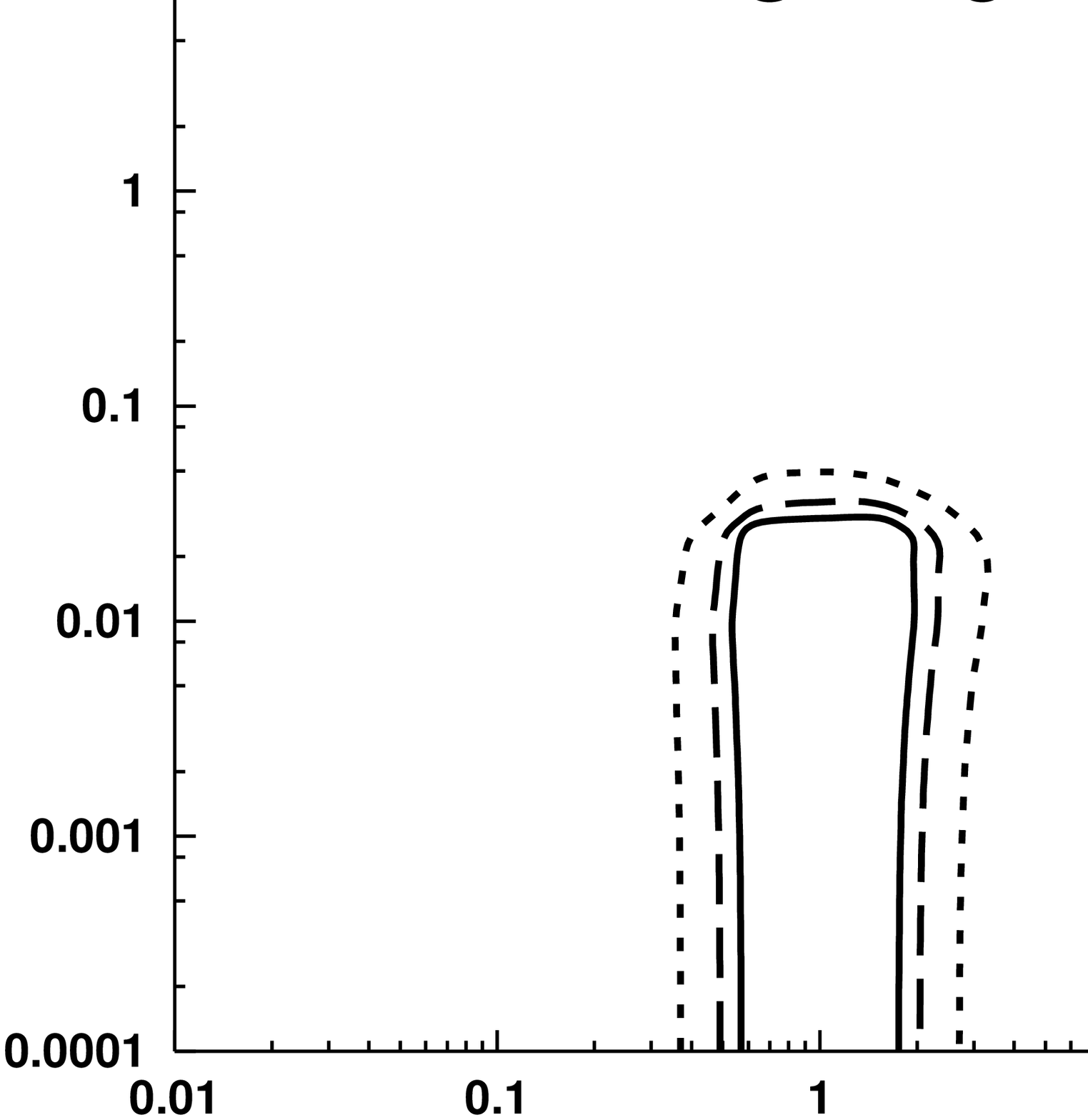,width=3.9cm}

\vglue 1.8cm
\hglue -1cm 
\epsfig{file=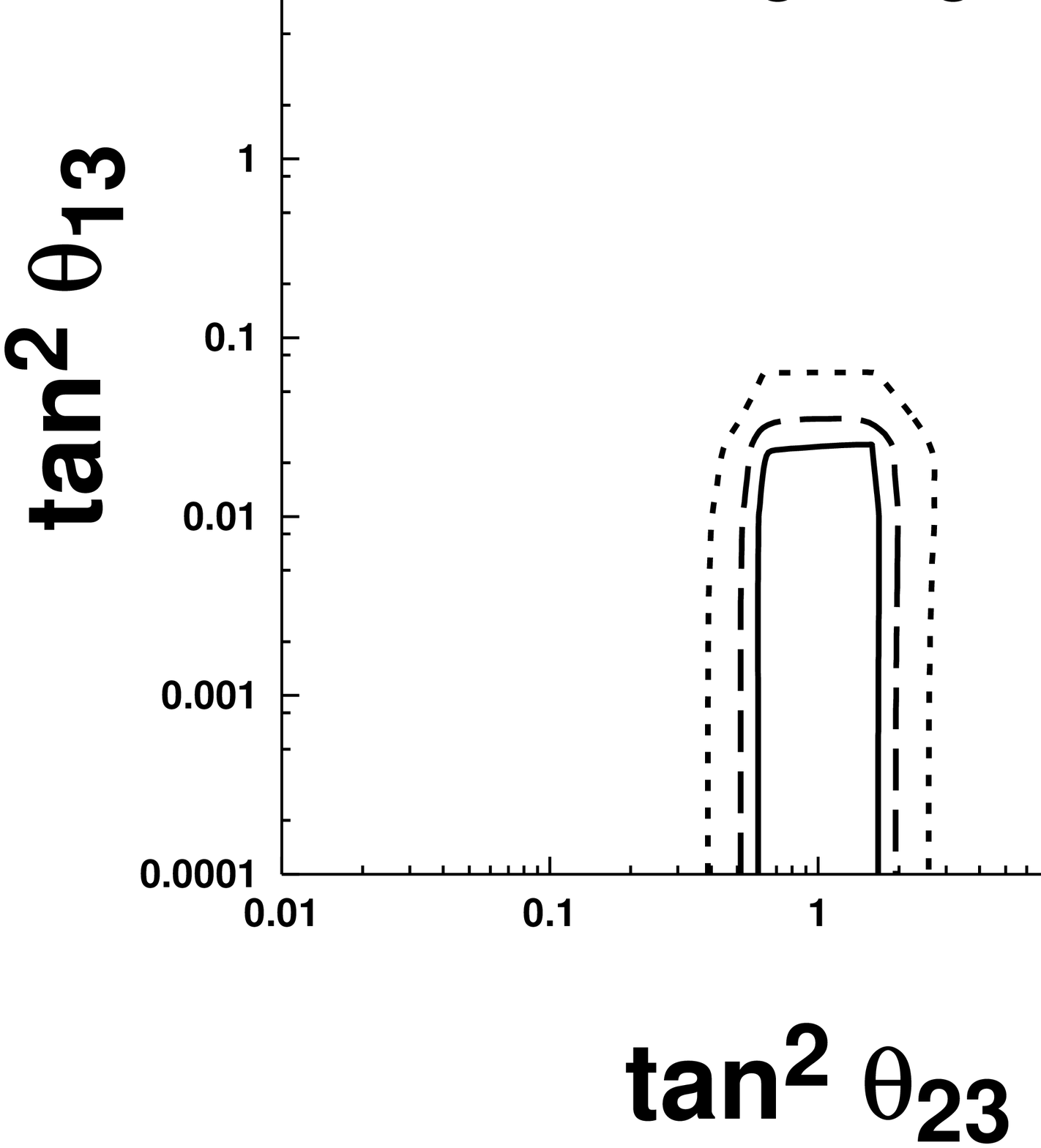,width=3.9cm}
\vglue -3.43cm \hglue 4.3cm \epsfig{file=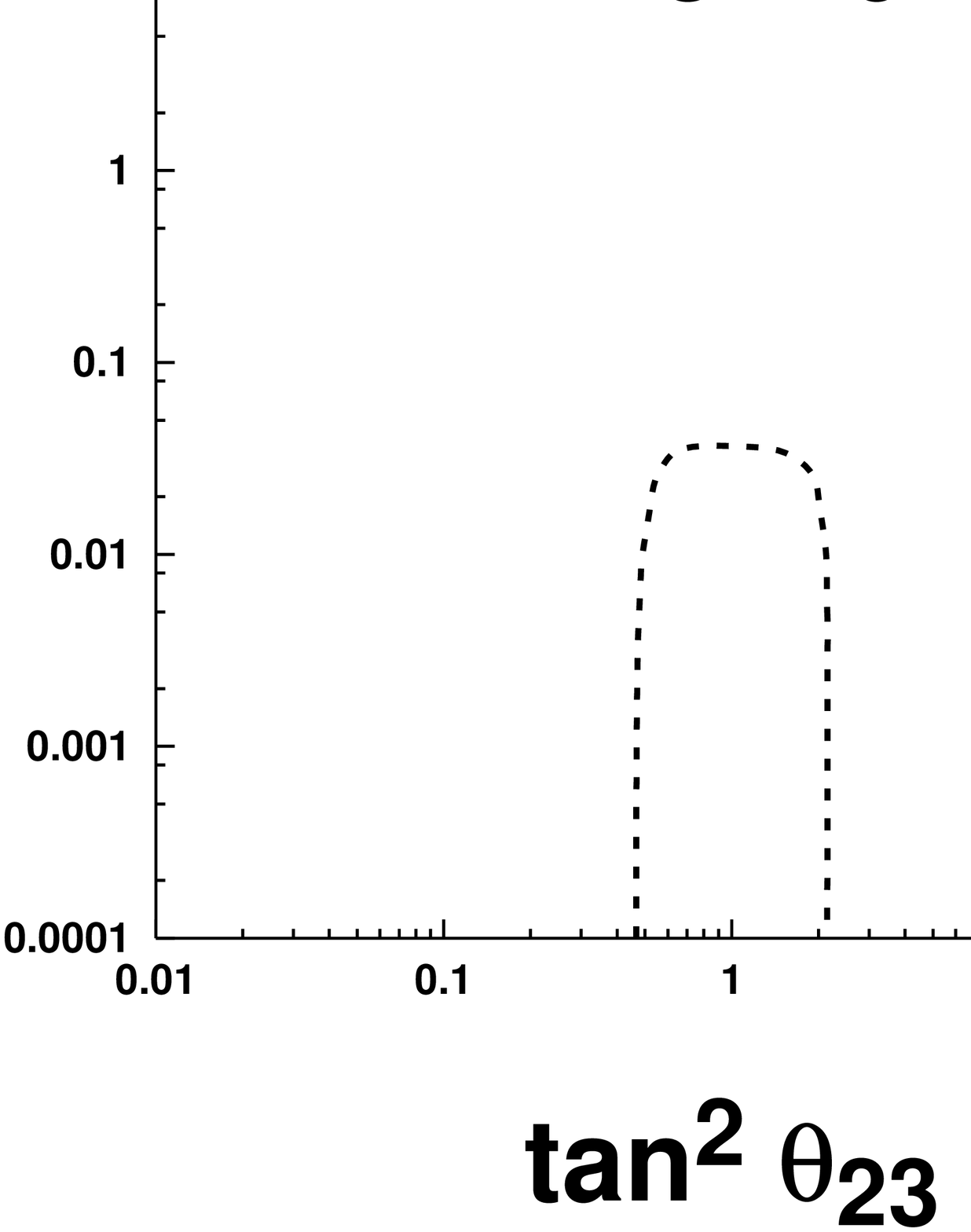,width=3.9cm}
\vglue -3.43cm \hglue 9.6cm \epsfig{file=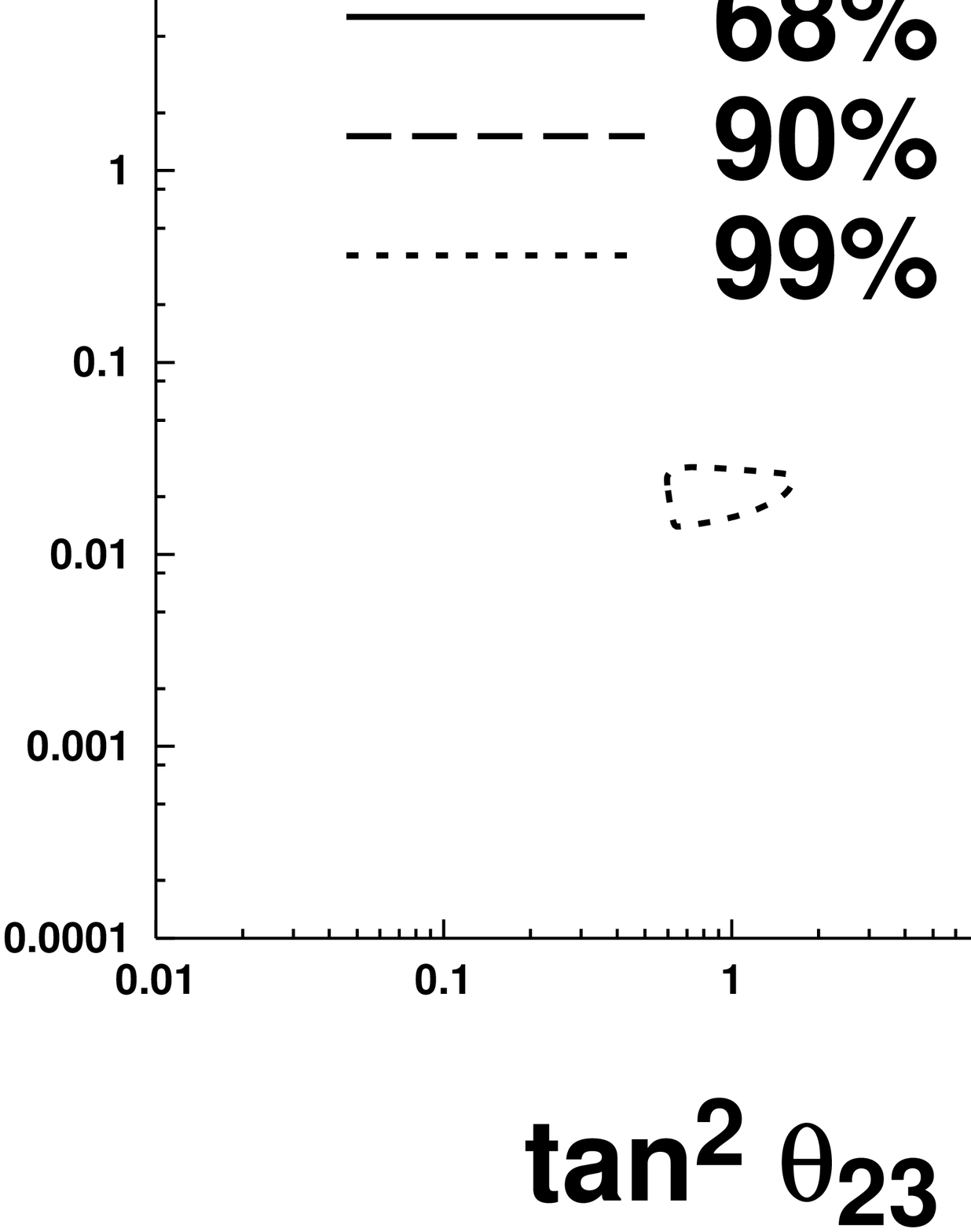,width=3.9cm}

\vglue -1.0cm
\hglue 6.0cm \epsfig{file=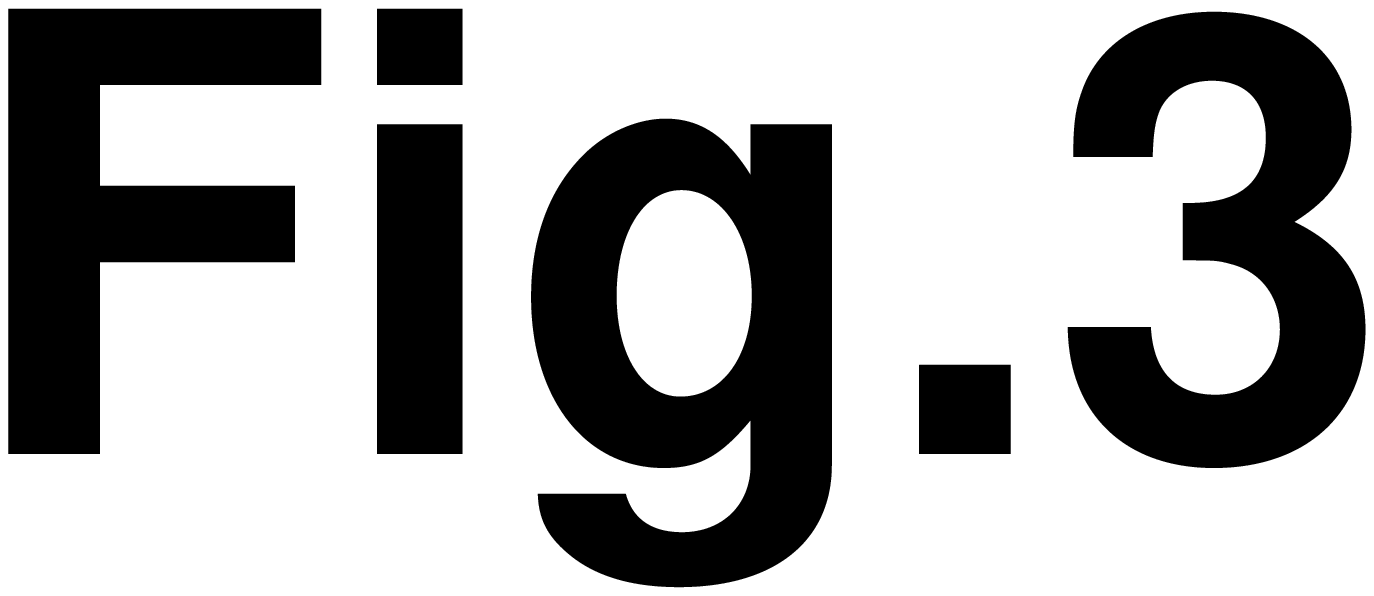,width=3.9cm}
\newpage
\epsfig{file=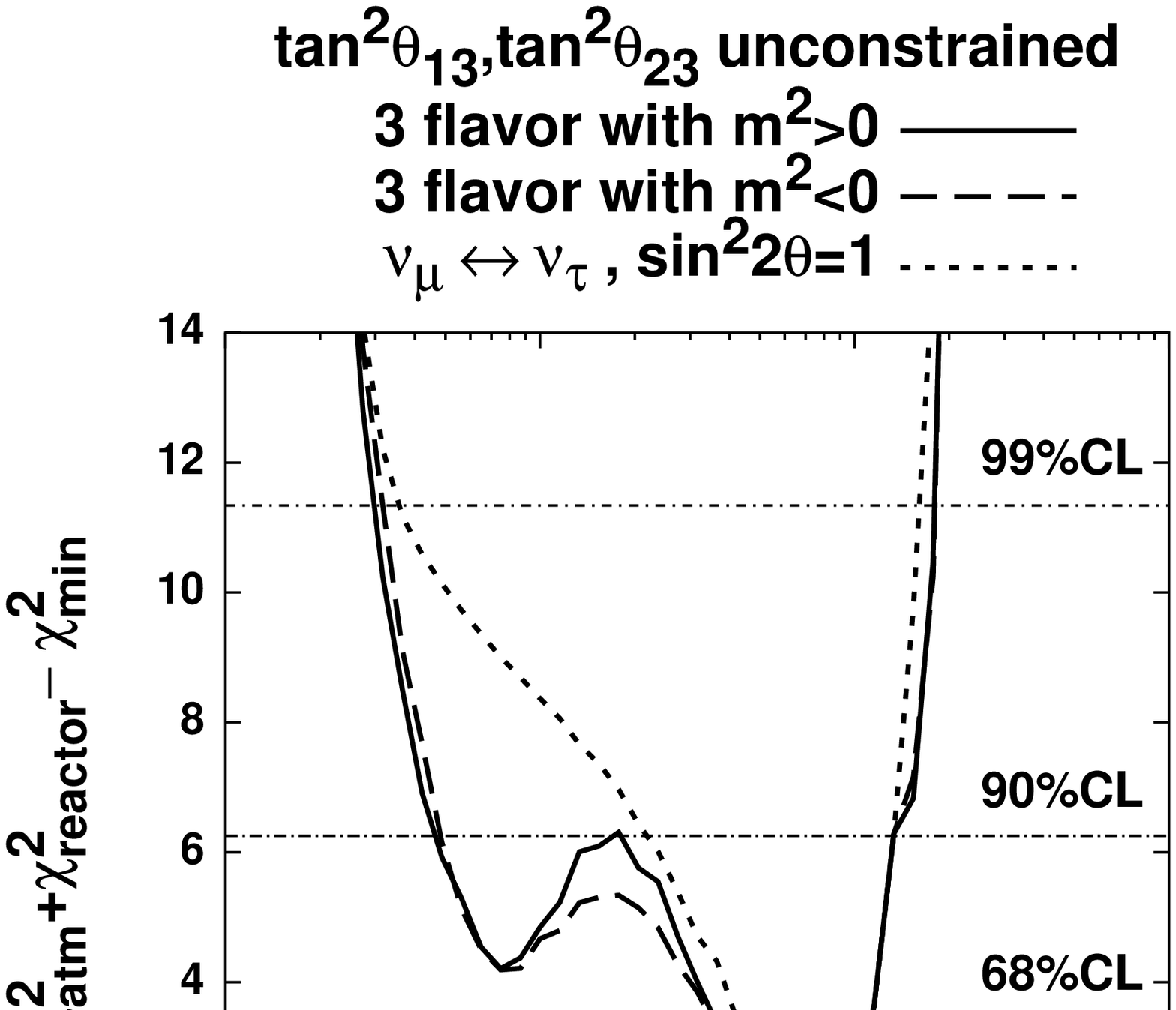,width=15cm}

\begin{thebibliography}{99}
\bibitem{largem}
If we lift the reactor constraint, the large $m^2$ limit
gives a reasonable fit to both the Kamiokande and Superkamiokande
data.  This is because the matter effect
$\sin^2\left(\int\sqrt{2}G_Fn_edx\right)$  can account for
the zenith angle dependence in case of three flavor oscillations
with large $\displaystyle\theta_{13}$\cite{flmm,fvy3}.
The author would like to thank E. Lisi for discussions on this point.
\bibitem{nb}
Notice that Figs. 2 and 3 do not have allowed region toward the right
side even for $| m^2 |<10^{-3}$ eV$^2$ where the reactor experiments
give no constraint.
\end{thebibliography}
\end{document}